\documentclass[preprintnumbers,article,amsmath,amssymb,floatfix,10pt,prd,onecolumn,
superscriptaddress,nofootinbib]{revtex4}
\usepackage{bbm}
\usepackage{amsfonts}
\usepackage{mathrsfs}
\usepackage{latexsym}

\usepackage{epsfig}
\usepackage{graphicx}
\usepackage{epstopdf}
\usepackage{placeins}
\usepackage{float}

\usepackage{amssymb}
\usepackage{amsmath}

\usepackage{dcolumn}
\usepackage{bm}
\usepackage{color}
\usepackage{comment}
\usepackage{xcolor}
\begin{document}
\title{\textbf{{Study of Compact Stars in $\mathcal{R}+ \alpha \mathcal{A}$ Gravity}}}

\author{M. Farasat Shamir}
\email{farasat.shamir@nu.edu.pk}\affiliation{National University of Computer and
Emerging Sciences,\\ Lahore Campus, Pakistan.}
\author{Eesha Meer}
\email{eeshameer129@gmail.com}\affiliation{National University of Computer and
Emerging Sciences,\\ Lahore Campus, Pakistan.}
\begin{abstract}

The main goal of this work is to provide a comprehensive study of relativistic structures in the context of recently proposed {$\mathcal{R}+ \alpha \mathcal{A}$} gravity, where $\mathcal{R}$ is the Ricci scalar, and $\mathcal{A}$ is the anti-curvature scalar. For this purpose, we examine a new classification of embedded class-I solutions of compact stars. To accomplish this goal, we consider an anisotropic matter distribution for {$\mathcal{R}+ \alpha \mathcal{A}$ gravity model} with static spherically symmetric spacetime distribution. Due to highly non-linear nature of field equations, we use the Karmarkar condition to link the $g_{rr}$ and $g_{tt}$ components of the metric. Further, we compute the values of constant parameters using the observational data of different compact stars. It is worthy to mention here that we chose a set of twelve important compact stars from the recent literature namely $4U~1538-52$, $SAX~J1808.4-3658$, $Her~X-1$, $LMC~X-4$, $SMC~X-4$, $4U~1820-30$, $Cen~X-3$, $4U~1608-52$, $PSR~J1903+327$, $PSR~J1614-2230$, $Vela~X-1$, $EXO~1785-248$. To evaluate the feasibility of {$\mathcal{R}+ \alpha \mathcal{A}$ gravity model}, we conduct several physical checks, such as evolution of energy density and pressure components, stability and equilibrium conditions, energy bounds, behavior of mass function and adiabatic index. It is concluded that {$\mathcal{R}+ \alpha \mathcal{A}$} gravity supports the existence of compact objects which follow observable patterns.
\end{abstract}

\maketitle

\date{\today}
\section{Introduction}
Investigation of relativistic stellar structures has now become an interesting research pursuit in astrophysics. The fundamental features of compact stars have encouraged several researchers not only in the prospect of usual general relativity ($\mathcal{GR}$) but also in broadly emerging gravitational alternative theories during the last few decades \cite{Skm, Bhr, Akp, SMR,  Vum, Bmk, Hin, Nag, Hat, Add}.

Compact stars are originated as the result of gravitational collapse that happens due to the pressure of the relativistic objects. The gravitational collapse occurs at the center of the massive star which is highly exothermic and this happens when the internal pressure of the stars fails to maintain the pressure against the external force. As the result of gravitational collapse, either a compact star will be born, like neutron stars, white dwarfs, or it will be extended as a black hole. $\mathcal{GR}$ provides most suitable results in the study of stellar objects and develop basic understanding about gravitational theories. However, $\mathcal{GR}$ alone does not provide satisfactory outcomes in order to determine the mysterious nature of dark energy. Due to the limitation of $\mathcal{GR}$, the modified theories have gained the attention of researchers. As an alternative to the theory of $\mathcal{GR}$, modified theories of gravity have played an important role to somehow explain the accelerating expansion of the universe. Various modified gravitational theories are presented in literature, some of mostly debated are $f(\mathcal{R})$, $f(\mathcal{R},\mathcal{T})$, $f(\mathcal{G})$, $f(\mathcal{G},\mathcal{T})$, $f(\mathcal{R},\mathcal{G})$ \cite{Hbl, MFS, Stn, Sma, Ssz}. A recently proposed theory namely $f(\mathcal{R}, \mathcal{A})$ gravity has also gained some popularity due to the involvement of bivariate function of Ricci and anti-curvature scalars \cite{Luca}.

We require a precise solution to the Einstein field equations ($\mathcal{EFE}$) to examine celestial compact structure models. Schwarzschild \cite{Sch} was the first who discovered the $\mathcal{EFE}$ solution for the internal structure of compact stars in 1916. Tolman \cite{Tol} and Oppenheimer \cite{Opp} explored some feasible models of stellar objects to represent the relationship between internal pressure and the gravitational force. Bowers and Liang \cite{Bal} introduced the idea of non-zero anisotropy in a stellar arrangement. Ruderman \cite{Rud} was the first to discover that the nuclear density becomes anisotropic at the center of compact objects. Errehymy et al. \cite{Err} investigated the solutions of dense stars in the domain of $\mathcal{GR}$. A new class of solution describing an anisotropic star which is satisfying Karmarker\textsc{\char13}s condition was reported by Maurya et al. \cite{Sbs}. The discussion of compact structures in modified gravity also seems interesting. Nojiri and Odintsov \cite{Nao} discussed some viable and stable results of different models in the framework of $f(\mathcal{R})$ gravity. Starobinsky \cite{Star} also presented the outcomes of various models of $f(\mathcal{R})$ gravity. Hu and Sawicki \cite{Has} suggested several $f(\mathcal{R})$ gravity models. Cognola et al. \cite{Cog} examined the physical behavior of the compact stars in the exponential type models of $f(\mathcal{R})$ gravity. Schlai \cite{Sli} identified the embedding problem on geometrically significant spacetimes. Shamir et al. \cite{Mir} explored the solutions of dense stars using Karmarkar condition in the background of $f(\mathcal{R})$ theory of gravity. Naz et al. \cite{Naz} investigated the existence of solutions of anisotropic compact objects through embedding approach in the context of $f(\mathcal{R})$ gravity by applying Karmarkar condition.

Motivated from the interesting properties of stellar structures in modified theories of gravity, we aim to investigate anisotropic compact structures in the background of Ricci inverse ($\mathcal{RI}$) gravity. In particular, we provide a detailed investigation of stellar structure for twelve important compact stars from the recent literature namely $4U~1538-52$, $SAX~J1808.4-3658$, $Her~X-1$, $LMC~X-4$, $SMC~X-4$, $4U~1820-30$, $Cen~X-3$, $4U~1608-52$, $PSR~J1903+327$, $PSR~J1614-2230$, $Vela~X-1$, $EXO~1785-248$. Our work is managed as follows: Section II presents the fundamental concepts, structure, and the modified field equations of $\mathcal{RI}$ gravity. In Section III, the matching constraints for the chosen {linear} $\mathcal{RI}$ gravity model ({$\mathcal{R}+ \alpha \mathcal{A}$}) are developed using Schwarzschild's geometry. Section IV is devoted to calculate and analyze physical properties such as energy density, pressure profiles etc. The last section contains final remarks.

\section{$f(\mathcal{R},\mathcal{A})$ Gravity and Modified Field Equations}

The inverse of the Ricci tensor known as the anticurvature tensor helps to form an alternative theory namely $f(\mathcal{R},\mathcal{A})$ gravity \cite{Luca},  where $\mathcal{R}$ and $\mathcal{A}$ are the traces of curvature and anticurvature respectively. This theory can be explored to verify the widespread presence of spectral instabilities and to understand the effects of perturbations such as the propagation of gravitational waves, perturbation growth, or the Newtonian bound. Such theory is confirmed as a suitable choice to be regarded as a cosmological model after deriving the general field equations of the Lagrangian $f(\mathcal{R},\mathcal{A})$. According to a no-go theorem, every Lagrangian action $f(\mathcal{R},\mathcal{A})$ containing terms with any positive or negative exponent of anticurvature $\mathcal{A}$ can be used to justify dark energy \cite{Luca}. Thus, it would be exceptionally intriguing to analyze this theory further. In this work, we focus ourselves to study compact structures in this theory. The starting point to derive the field equations is to define a tensor $\mathcal{A}^{\eta \xi}$ as inverse of $\mathcal{R}_{\eta \xi}$, such that
\begin{equation}\label{1}
\mathcal{A}^{\eta \alpha} \mathcal{R}_{\alpha \xi}= \delta^{\eta}_{\xi}.
\end{equation}
In order to develop the modified field equations, we need the following definition of the $f(\mathcal{R},\mathcal{A})$ gravitational action:
\begin{equation}\label{2}
S= \int d^4x\sqrt{-g}[f(\mathcal{R},\mathcal{A}) + \mathcal{L}_{m}],
\end{equation}
where, the arbitrary function $f(\mathcal{R},\mathcal{A})$ depends on the Ricci scalar $\mathcal{R}$ and the anticurvature scalar $\mathcal{A}$, which are the traces of the Ricci tensor and the anticurvature tensor respectively. Here, $\mathcal{L}_{m}$ is matter Lagrangian and $g$ represents the determinant of the metric. We get the following $f(\mathcal{R},\mathcal{A})$ gravity field equation by varying the action mentioned in Eq. (\ref{2}) with respect to the metric tensor
\begin{align}\label{3}
f_{\mathcal{R}} \mathcal{R}^{\eta \xi} - f_{\mathcal{A}} \mathcal{A}^{\eta \xi} - \frac{1}{2} f g^{\eta \xi} & + g^{\mu \eta} \nabla_{\beta} \nabla_{\mu} (\ f_{\mathcal{A}} \mathcal{A}_{\sigma}^{\beta} \mathcal{A}^{\xi \sigma})\ - \frac{1}{2} \nabla^{\kappa} \nabla_{\kappa} (\ f_{\mathcal{A}} \mathcal{A}_{\sigma}^{\eta} \mathcal{A}^{\xi \sigma})\ \nonumber \\& - \frac{1}{2} g^{\eta \xi} \nabla_{\beta}\nabla_{\mu} (\ f_{\mathcal{A}} \mathcal{A}_{\sigma}^{\beta} A^{\mu \sigma})\ - \nabla^{\eta} \nabla^{\xi} f_{\mathcal{R}} +  g^{\eta \xi} \nabla^{\kappa} \nabla_{\kappa}f_{\mathcal{R}} = \mathcal{T}^{\eta \xi},
\end{align}
where, $f = f(\mathcal{R},\mathcal{A})$, $f_{\mathcal{R}}= \frac{\partial f}{\partial \mathcal{R}}$, $f_{\mathcal{A}}= \frac{\partial f}{\partial \mathcal{A}}$, $\mathcal{A}^{\eta \xi}$ is the anticurvature tensor which is the inverse of the Ricci tensor $\mathcal{R}_{\eta \xi}$.
The stress-energy momentum tensor $\mathcal{T}_{\eta \xi}$ is chosen as
\begin{equation}\label{4}
\mathcal{T}_{\eta \xi}=(\rho + p_{t}) u_{\eta} u_{ \xi} - p_{t} g_{\eta \xi} + (p_{r} - p_{t}) \mathcal{\chi}_{\eta} \mathcal{\chi}_{\xi}.
\end{equation}
Here, $\rho,$ $p_{r}$ and $p_{t}$ are energy density, radial and transverse pressure components respectively, while $u_{\eta}$ and $\mathcal{\chi}_{\eta}$ are four velocity vectors, which satisfy the relations $u^{\eta} u_{\eta}= - \mathcal{\chi}^{\eta} \mathcal{\chi}_{\eta}=1$.
Furthermore, we consider a static spherically symmetric line element which is represented as
\begin{equation}\label{5}
ds^{2}=e^{\nu}dt^{2} - e^{\lambda}dr^{2} - r^{2}(d\theta^{2} + sin^{2}\theta d\phi^{2}),
\end{equation}
where, $\nu$ and $\lambda$ are the functions of radial coordinate. Our stress-energy tensor is assumed to be anisotropic, with relativistic geometrized units $8\pi G = 1$.
Moreover, the embedded class-I family is represented by the metric tensor (\ref{5}), if it fulfills the Karmarkar condition, i.e.
\begin{equation}\label{6}
\mathcal{R}_{1414} \mathcal{R}_{2323} = \mathcal{R}_{1212} \mathcal{R}_{3434} + \mathcal{R}_{1224} \mathcal{R}_{1334},
\end{equation}
with $\mathcal{R}_{2323}\neq 0$. The following differential equation for the given metric (\ref{5}) is developed by using the well-known Karmarkar condition as
\begin{equation}\label{7}
\lambda^{\prime} \nu^{\prime} - 2 \nu^{\prime \prime} - (\nu^{\prime})^{2} = \frac{\lambda^{\prime} \nu^{\prime}}{1 - e^{\lambda}},
\end{equation}
where $e^{\lambda} \neq 1$. Integrating Eq. (\ref{7}), we obtain
\begin{equation}\label{8}
e^{\nu} = \left[(A + B \int \sqrt{e^{\lambda} - 1} dr)^{2}\right].
\end{equation}
In this case, $A$ and $B$ are integration constants. Further, due to the highly non-linear nature of modified field equations, we have no other choice but to choose the component $g_{rr} = e^{\lambda(r)}$ in the following shape \cite{Bhr,Spt}
\begin{equation}\label{9}
e^{\lambda} =1 + a^{2} r^{2} (1 + b r^{2})^{x},
\end{equation}
where one can choose any real value of $x$ other than zero. It is really important to mention here that in order to establish embedded class-I spacetime, the parameters $a$ and $b$ should not be zero. In the domain of $\mathcal{GR}$, Singh and Pant \cite{Spt} investigated the properties of stellar structures by considering some positive values of $x$ for embedding class-I solutions of different models of compact stars. Later, Bhar et al. \cite{Bhr} also used this metric potential by taking the value of $x = - 4$, and the corresponding results are found stable in the context of $\mathcal{GR}$. Inspired by the work of Bhar et al. \cite{Bhr}, we aim to expand the study in the framework of {$\mathcal{RI}$ gravity} by choosing $x = - 4$. Interestingly, we tried some other possibilities for the choice of $x$ (both positive and negative), however, we could not obtain physically acceptable results in those cases. Manipulating Eqs. (\ref{8}) and (\ref{9}), the metric potential expression becomes
\begin{equation}\label{10}
e^{\nu} = \left[\left(A - \frac{a B}{2 b (1 + b r^{2})}\right)^{2}\right].
\end{equation}
\begin{figure} [H] \center
\begin{tabular}{cccc}
\epsfig{file=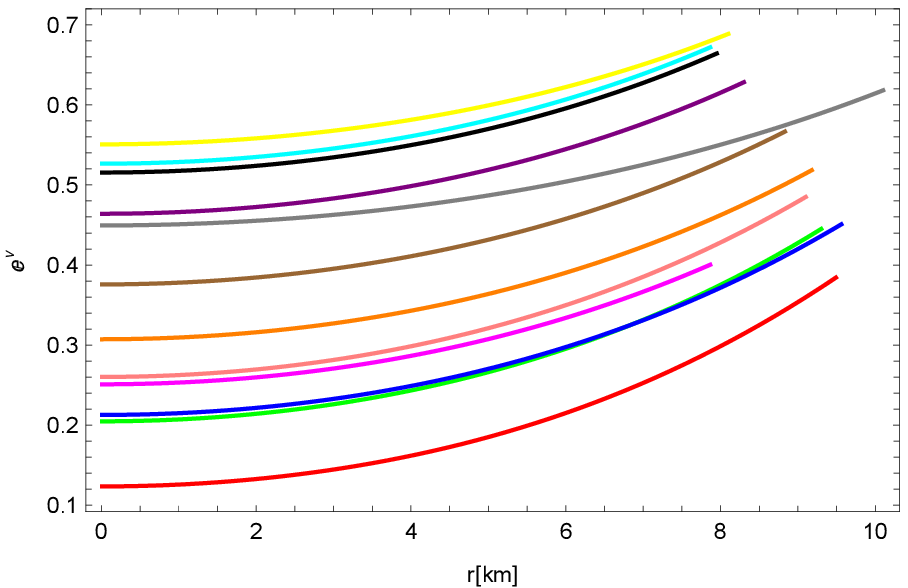,width=0.5\linewidth}  &
\epsfig{file=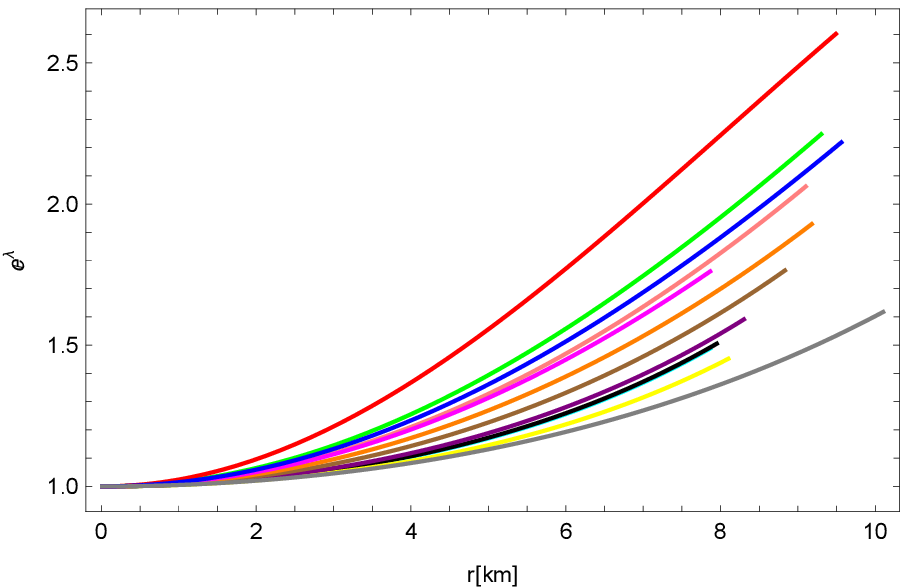,width=0.5\linewidth}  &
\end{tabular}
\caption{Graphical Behavior of $g_{tt}$ and $g_{rr}$ of $S_{1}$ ({\color{cyan} $\bigstar$}), $S_{2}$ ({\color{black} $\bigstar$}), $S_{3}$ ({\color{yellow} $\bigstar$}), $S_{4}$ ({\color{violet} $\bigstar$}), $S_{5}$ ({\color{brown} $\bigstar$}), $S_{6}$ ({\color{pink} $\bigstar$}), $S_{7}$ ({\color{orange} $\bigstar$}), $S_{8}$ ({\color{green} $\bigstar$}), $S_{9}$ ({\color{magenta} $\bigstar$}), $S_{10}$ ({\color{red} $\bigstar$}), $S_{11}$ ({\color{blue} $\bigstar$}), $S_{12}$ ({\color{gray} $\bigstar$}).}\center
\label{Fig:1}
\end{figure}
Further, to analyze the properties of compact stars, we probe the necessary conditions for the metric potentials of static spherically symmetric spacetime, namely $e^{\lambda(0)} = 1$ and $((e^{\lambda(r)})^{\prime})_{r=0} = 0$. The graphical response of $g_{rr}$ should not contain any singularities, and the curvature must be regular. From Fig. 1, it is evident that all above properties are satisfied and also the plot of $e^{\lambda}$ exhibits monotonically increasing behavior, reaching its maximum value near the boundary.

\section{Viable $\mathcal{RI}$ gravity model and matching conditions}

In this section, we provide some discussions about {linear} {$\mathcal{R}+ \alpha \mathcal{A}$ gravity model} and the matching conditions. By substituting this model in Eq.(\ref{3}), the field equations are simplified as
\begin{equation}\label{12}
\mathcal{R}^{\eta \xi} - \frac{1}{2} \mathcal{R} g^{\eta \xi} - \alpha \mathcal{A}^{\eta \xi} - \frac{1}{2} \alpha \mathcal{A} g^{\eta \xi} + \frac{\alpha}{2} (2 g^{\mu \eta} \nabla_{\beta}\nabla_{\mu} \mathcal{A}_{\sigma}^{\beta} \mathcal{A}^{\xi \sigma} - \nabla^{\kappa} \nabla_{\kappa} \mathcal{A}_{\sigma}^{\eta} \mathcal{A}^{\xi \sigma} - g^{\eta \xi} \nabla_{\beta}\nabla_{\mu} \mathcal{A}_{\sigma}^{\beta} \mathcal{A}^{\mu \sigma}) = \mathcal{T}^{\eta \xi}
\end{equation}
Using Eq. (\ref{12}) and spacetime (\ref{5}), modified field equations turn out to be
\begin{align}\label{13}
\rho &= e^{- \nu} \Gamma_{1} - \frac{1}{2} \Gamma_{2} - \alpha e^{\nu} \Gamma_{3} - \frac{1}{2} \alpha \Gamma_{4} + \frac{\alpha}{2} (e^{\nu - \lambda} \Gamma_{5}^{\prime \prime} + e^{\nu - \lambda} \nu^{\prime \prime} \Gamma_{5} +  2 e^{\nu - \lambda} \nu^{\prime} \Gamma_{5}^{\prime}  + e^{\nu - \lambda} (\nu^{\prime})^{2} \Gamma_{5} - \frac{\lambda^{\prime}}{2} e^{\nu - \lambda} \Gamma_{5}^{\prime} \nonumber\\& - \frac{\nu^{\prime}\lambda^{\prime}}{2} e^{\nu - \lambda} \Gamma_{5} + \frac{2}{r} e^{\nu - \lambda} \Gamma_{5}^{\prime} + \frac{2 \nu^{\prime}}{r} e^{\nu - \lambda} \Gamma_{5} - \Gamma_{6}^{\prime \prime} - \lambda^{\prime \prime} \Gamma_{6} - \frac{3}{2} \lambda^{\prime} \Gamma_{6}^{\prime} - \frac{(\lambda^{\prime})^{2}}{2} \Gamma_{6} - 2 r e^{- \lambda} \Gamma_{7}^{\prime} - \frac{4}{r^{2}} \Gamma_{6}),
\end{align}

\begin{align}\label{14}
p_{r} &= e^{- \lambda} \Gamma_{8} + \frac{1}{2} \Gamma_{2} - \alpha e^{\lambda} \Gamma_{9} + \frac{1}{2} \alpha \Gamma_{4} + \frac{\alpha}{2} (- \frac{(\nu^{\prime})^{2}}{2} e^{\nu - \lambda} \Gamma_{5} + \frac{\nu^{\prime} \lambda^{\prime}}{2} \Gamma_{6} + \frac{\nu^{\prime}}{2} \Gamma_{6}^{\prime} + 4 e^{- \lambda} \Gamma_{7}+ \frac{2 \lambda^{\prime}}{r} \Gamma_{6} + \frac{2}{r} \Gamma_{6}^{\prime} \nonumber\\& - \frac{\nu^{\prime}}{2} e^{\nu - \lambda} \Gamma_{5}^{\prime} + 2 r e^{- \lambda} \Gamma_{7}^{\prime}),
\end{align}

\begin{align}\label{15}
p_{t} &= \frac{1}{r^{2}} \Gamma_{10} + \frac{1}{2} \Gamma_{2} - \alpha r^{2} \Gamma_{11} + \frac{1}{2} \alpha \Gamma_{4} + \frac{\alpha}{2} (\frac{\nu^{\prime}}{2} r^{2} e^{- \lambda} \Gamma_{7}^{\prime} + r e^{- \lambda} \nu^{\prime} \Gamma_{7} + r^{2} e^{- \lambda} \Gamma_{7}^{\prime \prime} + 4 e^{- \lambda} \Gamma_{7} + 6 r e^{- \lambda} \Gamma_{7}^{\prime} \nonumber\\& - \frac{\lambda^{\prime}}{2} r^{2} e^{- \lambda} \Gamma_{7}^{\prime} - r e^{- \lambda} \lambda^{\prime} \Gamma_{7} - \frac{\nu^{\prime}}{2} e^{\nu - \lambda} \Gamma_{5}^{\prime} + \frac{(\nu^{\prime})^{2}}{2} \Gamma_{6} + \Gamma_{6}^{\prime \prime} + \lambda^{\prime \prime} \Gamma_{6} + \frac{3}{2} \lambda^{\prime} \Gamma_{6}^{\prime} + \frac{(\lambda^{\prime})^{2}}{2} \Gamma_{6} + \frac{2}{r^{2}} \Gamma_{6}),
\end{align}
where, the values of $\Gamma_{i}$'s $(i=1...11)$ are mentioned in Appendix and the prime represents the derivative with respect to $r$.

The star's geometrical structure, whether viewed from the outside or the inside, has no impact on the interior boundary metric. The metric components must be continuous to the boundary regardless of the referential frame in this emerging situation. While evaluating several matching conditions and exploring compact objects in the context of $\mathcal{GR}$, the Schwarzschild solution seems to be the right approach. According to the Jebsen-Birkhoff theorem, given spherically symmetric spacetime, the solution of the Einstein field equations $\mathcal{(EFE)}$ must be asymptotically flat and static. In modified gravitational theories, the exterior geometry solution may vary from the Schwarzschild solution when we consider modified $\mathcal{(TOV)}$ equations \cite{Tol, Opp} with zero pressure and energy density. Furthermore, the Schwarzschild solution can be achieved in $\mathcal{RI}$ gravity by choosing a realistic and viable {$\mathcal{R}+ \alpha \mathcal{A}$ gravity model} with non-zero density and pressure components. The Birkhoff theorem contradicts modified theories as a result \cite{Vfi}. Numerous studies using Schwarzschild solution in matching constraints have produced impressive findings \cite{Ava, Acn, Agy, Dmi}. The radial pressure $p_{r}(r = R) = 0$ is used to calculate the field equations solution under the specified boundary conditions at $r = R$. Thus, we compare the internal geometry solution to the Schwarzschild external geometry provided by
\begin{equation}\label{14}
ds^{2}=\left( 1-\frac{2M}{r}\right)dt^{2} - \left(1-\frac{2M}{r}\right)^{-1} dr^{2} - r^{2}(d\theta^{2} + sin^{2}\theta d\phi^{2}),
\end{equation}
where the entire mass contained within the star is represented by $M$. The following expressions are derived at $r = R$ by probing at the metric potentials as follows:
\begin{align}\label{15}
g_{tt}^{+} = g_{tt}^{-}, \qquad  g_{rr}^{+} = g_{rr}^{-}, \qquad  \frac{\partial g_{tt}^{+}}{\partial r} = \frac{\partial g_{tt}^{-}}{\partial r},
\end{align}
where, (+) refers to the external geometry and (-) represents the internal geometry. The values of the required unknown quantities can be obtained by comparing the inner and outer matrices as shown below:
\begin{align}\label{16,17}
a&=\frac{(1+bR^{2})^{2}}{R} \sqrt{\frac{\frac{2M}{R}}{1-\frac{2M}{R}}}, &  b&=\frac{4M-R}{R^{2}(9R-20M)},\\
A&=\frac{aB}{2b(1+bR^{2})}+\sqrt{1-\frac{2M}{R}}, &  B&=\left(\frac{1}{2R}\right) \sqrt{\frac{2M}{R}}.
\end{align}
We will use these matching conditions to analyze different compact star models in the background of {$\mathcal{R}+ \alpha \mathcal{A}$ gravity}. It is worthwhile to mention here that we provide a detailed comparison by considering $12$ compact star models namely $4U~1538-52$, $SAX~J1808.4-3658$, $Her~X-1$, $LMC~X-4$, $SMC~X-4$, $4U~1820-30$, $Cen~X-3$, $4U~1608-52$, $PSR~J1903+327$, $PSR~J1614-2230$, $Vela~X-1$, $EXO~1785-248$. {The values of the input parameters radius and mass of considered compact stars are given in Table I:}

\begin{table}[H]
\centering
\scalebox{1}{
 \begin{tabular}{|| c | c | c ||}
\hline
\textbf{Star Model} & $M (M_{\Theta})$ & $R (km)$ \\ [0.5ex]
\hline\hline
 \textbf{4U~1538-52} $(S_{1})$ & 0.87 $\pm$ 0.07 \cite{SiU} & 7.866 $\pm$ 0.21 \\
 \hline
 \textbf{SAX~J1808.4-3658} $(S_{2})$ & 0.9 $\pm$ 0.3 \cite{Ele} & 7.951 $\pm$ 1.0 \\
 \hline
 \textbf{Her~X-1} $(S_{3})$ & 0.85 $\pm$ 0.15 \cite{MKA} & 8.1 $\pm$ 0.41 \\
 \hline
 \textbf{LMC~X-4} $(S_{4})$ & 1.04 $\pm$ 0.09 \cite{SiU} & 8.301 $\pm$ 0.2 \\
 \hline
 \textbf{SMC~X-4} $(S_{5})$ & 1.29 $\pm$ 0.05 \cite{SiU} & 8.831 $\pm$ 0.09 \\
 \hline
 \textbf{4U~1820-30} $(S_{6})$ & 1.58 $\pm$ 0.06 \cite{GT} & 9.1 $\pm$ 0.4 \\
 \hline
 \textbf{Cen~X-3} $(S_{7})$ & 1.49 $\pm$ 0.08 \cite{SiU} & 9.178 $\pm$ 0.13 \\
 \hline
 \textbf{4U~1608-52} $(S_{8})$ & 1.74 $\pm$ 0.01 \cite{Wro} & 9.3 $\pm$ 0.10 \\
 \hline
 \textbf{PSR~J1903+327} $(S_{9})$ & 1.667 $\pm$ 0.021 \cite{PCC} & 9.48 $\pm$ 0.03 \\
 \hline
 \textbf{PSR~J1614-2230} $(S_{10})$ & 1.97 $\pm$ 0.04 \cite{DEM} & 9.69 $\pm$ 0.2 \\
 \hline
 \textbf{Vela~X-1} $(S_{11})$ & 1.77 $\pm$ 0.08 \cite{SiU} & 9.56 $\pm$ 0.08 \\
 \hline
 \textbf{EXO~1785-248} $(S_{12})$ & 1.30 $\pm$ 0.2 \cite{Oze}& 10.10 $\pm$ 0.44 \\ [1ex]
\hline
\end{tabular}}
\caption{Approximated values of input parameters}
\label{tab: Table 1}
\end{table}

{By using the values of the input parameters from Table I and manipulating Eqs. (\ref{16,17}), the values of the output parameters $a$, $b$, $A$, and $B$ are shown in the following Table:}

\begin{table}[H]
\centering
\scalebox{1}{
 \begin{tabular}{|| c | c | c | c | c ||}
\hline
\textbf{Star Model} & $a (km)$ & $b (km)$ & $A (km)$ & $B (km)$ \\ [0.5ex]
\hline\hline
 \textbf{4U~1538-52} $(S_{1})$ & 0.0784715 & -0.00097169 & -0.744456 & 0.0364083\\
 \hline
 \textbf{SAX~J1808.4-3658} $(S_{2})$ & 0.0793101 & -0.000920891 & -0.851084 & 0.0364385 \\
 \hline
 \textbf{Her~X-1} $(S_{3})$ & 0.0726951 & -0.000977202 & -0.538845 & 0.0344393\\
 \hline
 \textbf{LMC~X-4} $(S_{4})$ & 0.0838586 & -0.000705186 & -1.50207 & 0.0367192\\
 \hline
 \textbf{SMC~X-4} $(S_{5})$ & 0.0934356 & -0.000366545 & -4.13713 & 0.0372694\\
 \hline
 \textbf{4U~1820-30} $(S_{6})$ & 0.115014 & 0.0000941913 & 24.5842 & 0.039431\\
 \hline
 \textbf{Cen~X-3} $(S_{7})$ & 0.103164 & -0.000104682 & -18.0737 & 0.0378044\\
 \hline
 \textbf{4U~1608-52} $(S_{8})$ & 0.127851 & 0.000368416 & 7.40205 & 0.0400518\\
 \hline
 \textbf{PSR~J1903+327} $(S_{9})$ & 0.112672 & 0.000126983 & 17.4001 & 0.0380914\\
 \hline
 \textbf{PSR~J1614-2230} $(S_{10})$ & 0.156003 & 0.0009043 & 3.91745 & 0.0413433\\
 \hline
 \textbf{Vela~X-1} $(S_{11})$ & 0.122018 & 0.000306813 & 8.16831 & 0.0387589\\
 \hline
 \textbf{EXO~1785-248} $(S_{12})$ & 0.0708703 & -0.000447232 & -1.75321 & 0.0305887\\ [1ex]
\hline
\end{tabular}}
\caption{Approximated values of output parameters with $\alpha = 1.3\times10^{-11}$}
\label{tab: Table 2}
\end{table}

Further, the requirements for a well-behaved compact structure are specified as:\\\\
$\bullet$ The plots of $\rho$, $p_r$, and $p_t$ must always be positive, finite, and maximum at the center.\\
$\bullet$ The gradient of $\rho$, $p_r$, and $p_t$ should be negative.\\
$\bullet$ All of the energy requirements must be achieved.\\
$\bullet$ The equation of state ($EoS$) should satisfy the mandatory constraints, namely $0 < \omega_r$, $\omega_t < 1$.\\
$\bullet$ The equilibrium stability condition must be fulfilled by all of the forces.\\
$\bullet$ The parameters of speed of sound must be between $[0, 1]$, i.e. $0 < \upsilon_r^2$, and $\upsilon_t^2 < 1$.\\
$\bullet$ Adiabatic index must be greater than $\frac{4}{3}$ for an anisotropic fluid sphere.

\section{Physical characteristics of the compact star models}

In this section, we discuss some important physical properties of the observed compact stars in the context of considered {$\mathcal{R}+ \alpha \mathcal{A}$ gravity model}. For this purpose, we provide the graphical representations of energy density, pressure components, energy conditions, mass function, adiabatic index, stability and equilibrium conditions. Moreover, for our given model, we take the parameter $\alpha = 1.3\times10^{-11}$ for various compact stars. {The parameter $\alpha$ has a very limited range which provides us physically accepted results. We have checked two other values of parameter $\alpha = 1.0\times10^{-9}$, $\alpha = 1.9\times10^{-8}$ and found that there is a very slight difference in the behavior of the graphs towards the boundary. That difference is shown for one star $Vela~X-1$ as a magnified image in Fig. 2.}

\subsection{\textbf{Energy Density and Pressure Profiles}}

We display the graphical behavior of $\rho$, $p_r$ and $p_t$ for {$\mathcal{R}+ \alpha \mathcal{A}$ model} in this subsection. The graphical action of these aspects is presented in Fig. 2. In addition, we also show the graphical response of gradients of $\rho$, $p_r$ and $p_t$ in the Fig 3.
It is clearly seen from the Fig. 2 that plots of energy density, radial and transverse pressures are positive as well as finite. One can easily notice that the demonstration of $\rho$, $p_r$ and $p_t$ components reach the maximum values at the center of compact star and decline towards the stellar surface boundary.
\begin{figure} [H] \center
\begin{tabular}{cccc}
\epsfig{file=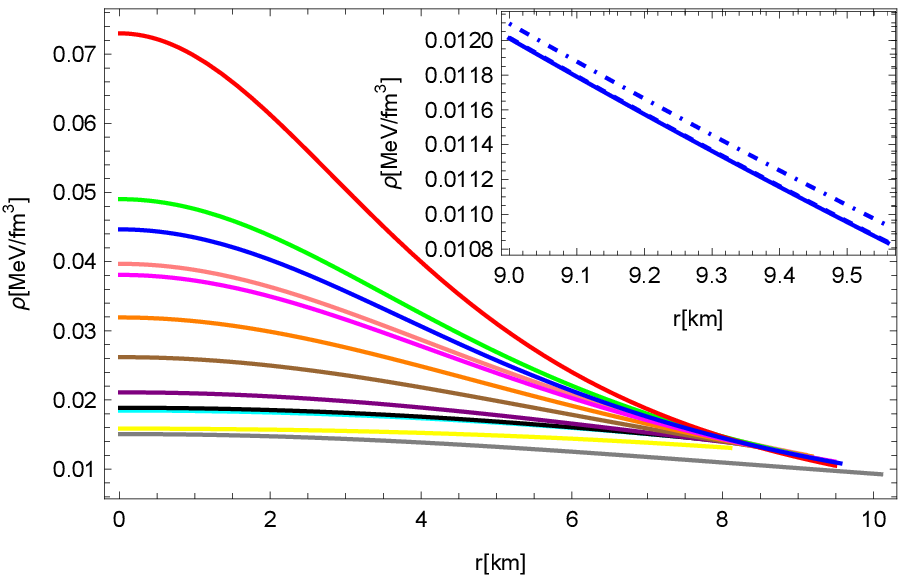,width=0.33\linewidth}  &
\epsfig{file=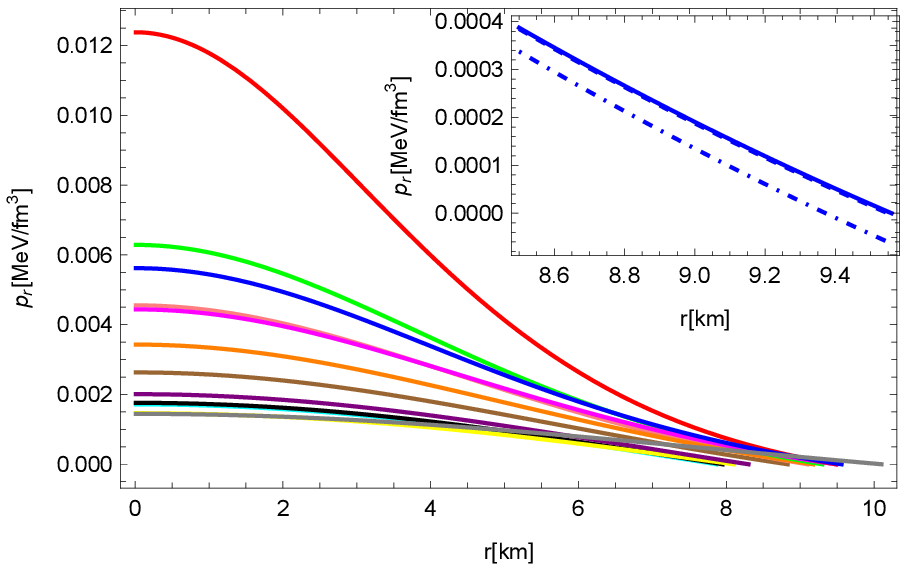,width=0.33\linewidth}  &
\epsfig{file=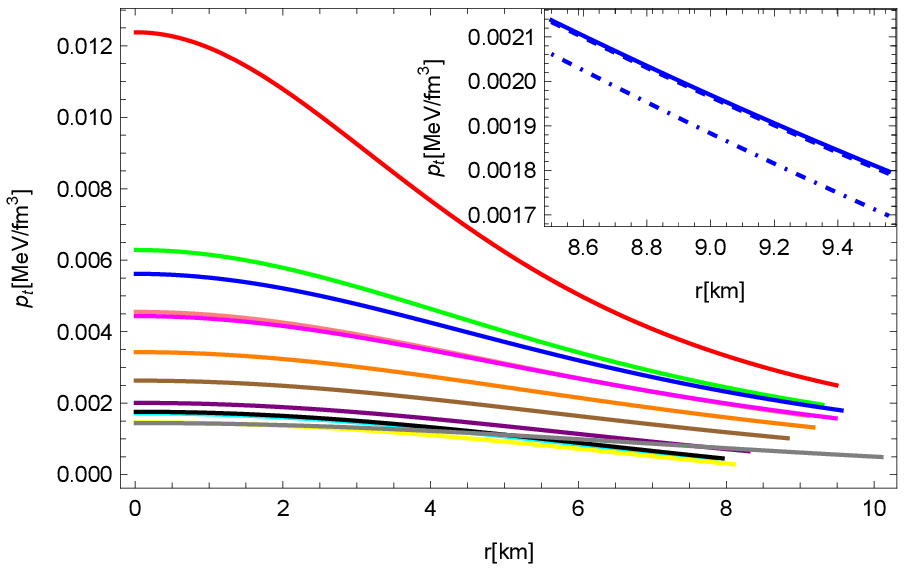,width=0.33\linewidth}  &
\end{tabular}
\caption{Behavior of $\rho$, $p_r$ and $p_t$ of $S_{1}$ ({\color{cyan} $\bigstar$}), $S_{2}$ ({\color{black} $\bigstar$}), $S_{3}$ ({\color{yellow} $\bigstar$}), $S_{4}$ ({\color{violet} $\bigstar$}), $S_{5}$ ({\color{brown} $\bigstar$}), $S_{6}$ ({\color{pink} $\bigstar$}), $S_{7}$ ({\color{orange} $\bigstar$}), $S_{8}$ ({\color{green} $\bigstar$}), $S_{9}$ ({\color{magenta} $\bigstar$}), $S_{10}$ ({\color{red} $\bigstar$}), $S_{11}$ ({\color{blue} $\bigstar$}), $S_{12}$ ({\color{gray} $\bigstar$}). {The magnified images in the figure show the behavior of the star $Vela~X-1$ for two different values of parameter $\alpha = 1.0\times10^{-9}$, $\alpha = 1.9\times10^{-8}$.}}\center
\label{Fig:2}
\end{figure}
In Fig. 3, the gradients of $\rho$, $p_r$ and $p_t$ are plotted which are finite but negative. These elements indicate the high compactness character of the compact stars.
\begin{figure} [H] \center
\begin{tabular}{cccc}
\epsfig{file=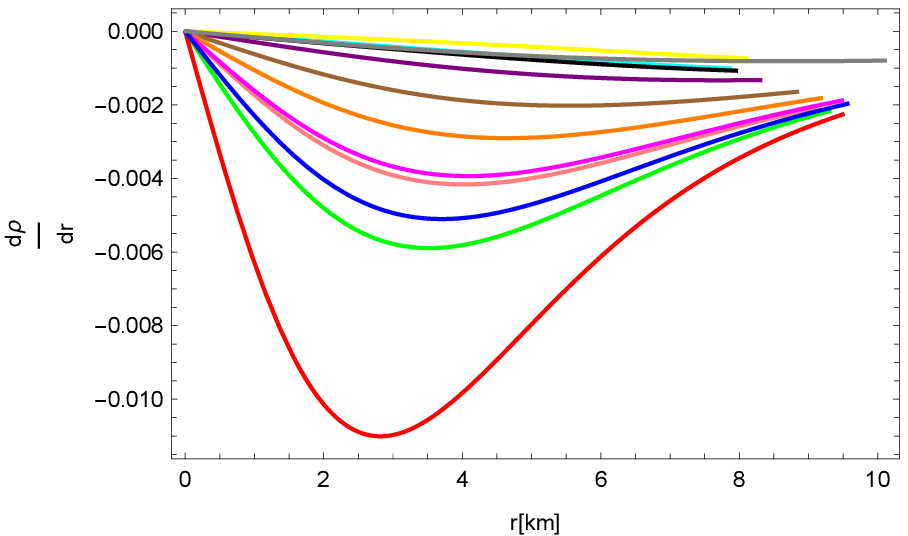,width=0.33\linewidth}  &
\epsfig{file=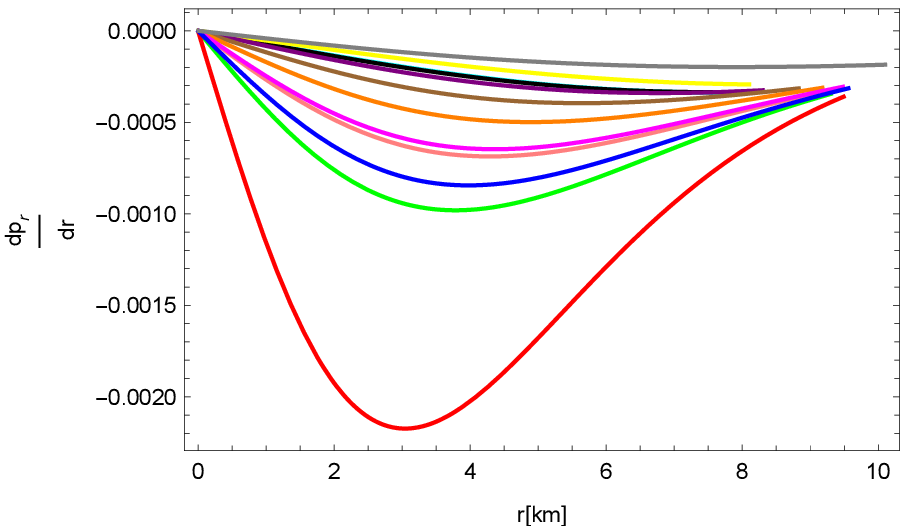,width=0.33\linewidth}  &
\epsfig{file=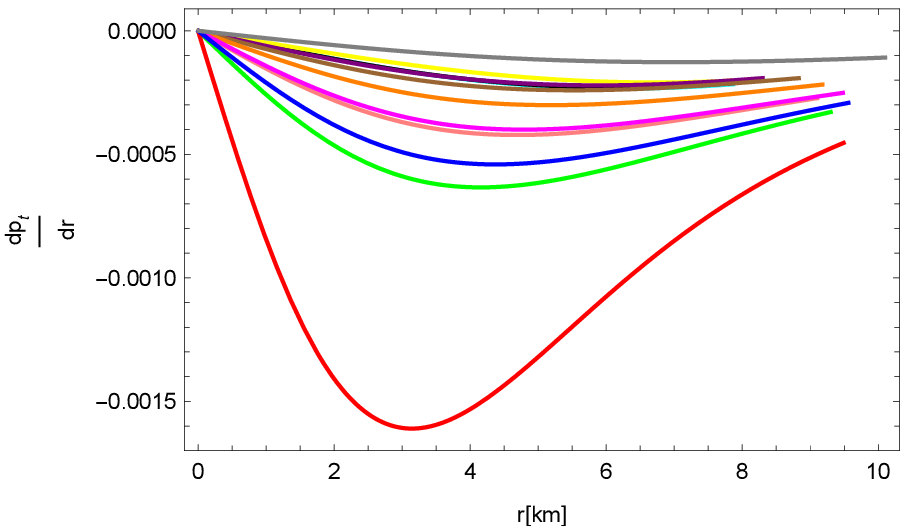,width=0.33\linewidth}  &
\end{tabular}
\caption{Variation of $\frac{d\rho}{dr}$, $\frac{dp_r}{dr}$ and $\frac{dp_t}{dr}$ of $S_{1}$ ({\color{cyan} $\bigstar$}), $S_{2}$ ({\color{black} $\bigstar$}), $S_{3}$ ({\color{yellow} $\bigstar$}), $S_{4}$ ({\color{violet} $\bigstar$}), $S_{5}$ ({\color{brown} $\bigstar$}), $S_{6}$ ({\color{pink} $\bigstar$}), $S_{7}$ ({\color{orange} $\bigstar$}), $S_{8}$ ({\color{green} $\bigstar$}), $S_{9}$ ({\color{magenta} $\bigstar$}), $S_{10}$ ({\color{red} $\bigstar$}), $S_{11}$ ({\color{blue} $\bigstar$}), $S_{12}$ ({\color{gray} $\bigstar$}).}\center
\label{Fig:3}
\end{figure}

\subsection{\textbf{Anisotropy}}

In this portion, we discuss the graphical response of anisotropy parameter \cite{Cgb} which is represented by $\Delta$ and defined as the difference between the transverse and radial pressures i.e. $\Delta = p_t - p_r$. The nature of $\Delta$ is attractive if $p_t < p_r$ and repulsive if $p_t > p_r$. Therefore, the anisotropy must indicate repulsive nature \cite{Mkg} for the compact stars to exist. From Fig. 4, it is obvious that anisotropy remains positive in our case i.e. $\Delta > 0$, which implies that anisotropic force is drawn outwards.
\begin{figure} [H] \center
\begin{tabular}{cccc}
\epsfig{file=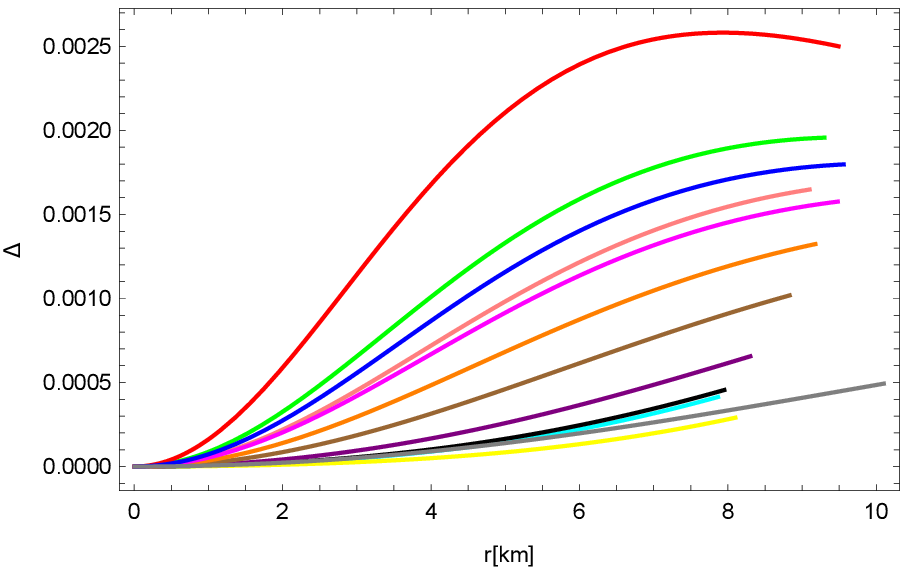,width=0.5\linewidth}  &
\end{tabular}
\caption{Evolution of $\Delta$ of $S_{1}$ ({\color{cyan} $\bigstar$}), $S_{2}$ ({\color{black} $\bigstar$}), $S_{3}$ ({\color{yellow} $\bigstar$}), $S_{4}$ ({\color{violet} $\bigstar$}), $S_{5}$ ({\color{brown} $\bigstar$}), $S_{6}$ ({\color{pink} $\bigstar$}), $S_{7}$ ({\color{orange} $\bigstar$}), $S_{8}$ ({\color{green} $\bigstar$}), $S_{9}$ ({\color{magenta} $\bigstar$}), $S_{10}$ ({\color{red} $\bigstar$}), $S_{11}$ ({\color{blue} $\bigstar$}), $S_{12}$ ({\color{gray} $\bigstar$}).}\center
\label{Fig:4}
\end{figure}

\subsection{\textbf{Energy Conditions}}

The significant role of energy conditions in describing the existence of anisotropic compact stars is widely known in literature \cite{Zya}. These energy limitations are classified namely null energy, strong energy, weak energy and dominant energy bounds represented by $\mathcal{NEC}$, $\mathcal{SEC}$, $\mathcal{WEC}$ and $\mathcal{DEC}$, sequentially and expressed as
\begin{align}
\mathcal{NEC}: \rho + p_r \ge 0, \quad  \rho + p_t \ge 0, \qquad \mathcal{WEC}: \rho \ge 0, \quad  \rho + p_r \ge 0, \quad \rho + p_t \ge 0, \nonumber \\
\mathcal{SEC}: \rho + p_r + 2p_t \ge 0, \qquad \mathcal{DEC}: \rho \ge 0, \quad \rho - |p_r| \ge 0, \quad \rho - |p_t| \ge 0.
\end{align}
From Fig. 5, it is clearly observed that all energy conditions are satisfied for considered {$\mathcal{R}+ \alpha \mathcal{A}$ gravity model}.
\begin{figure} [H] \center
\begin{tabular}{cccc}
\epsfig{file=rho.eps,width=0.33\linewidth}  &
\epsfig{file=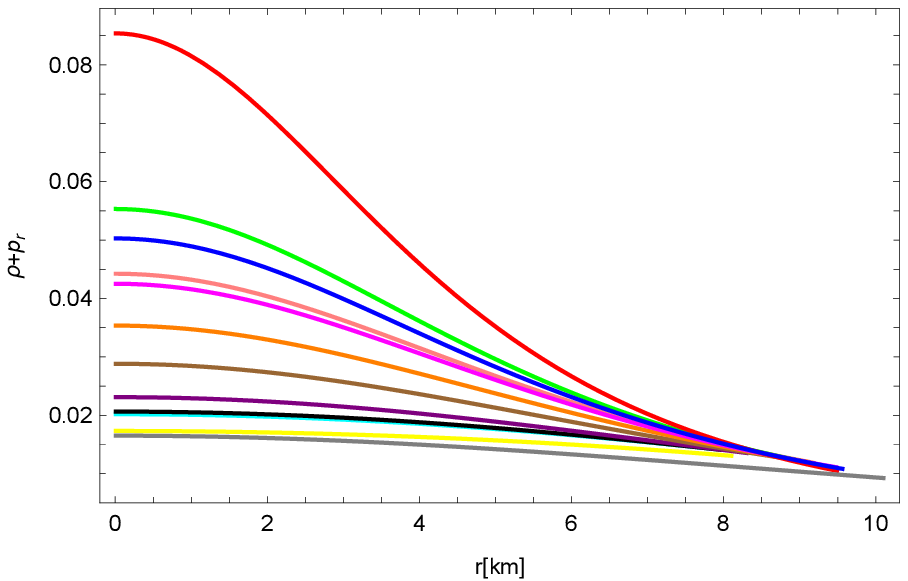,width=0.33\linewidth}  &
\epsfig{file=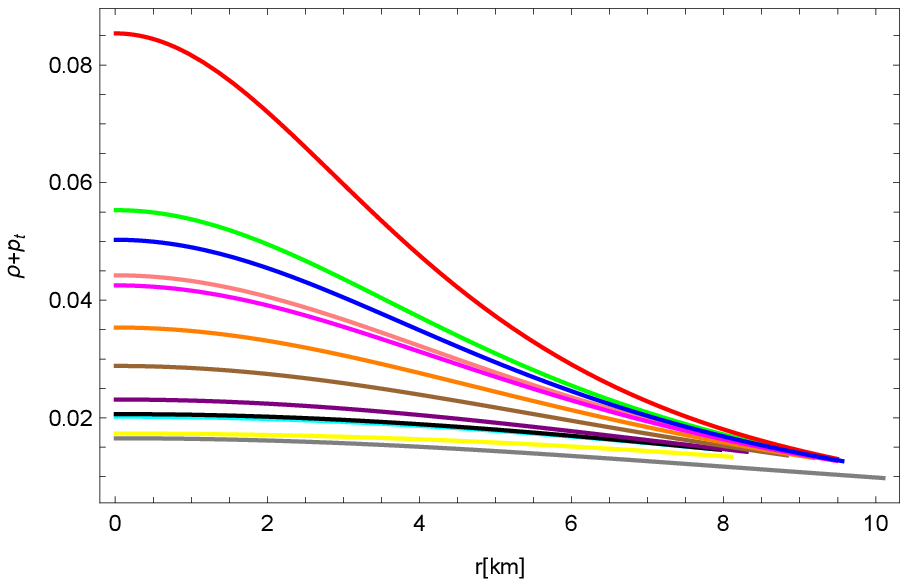,width=0.33\linewidth}  &
\end{tabular}
\begin{tabular}{cccc}
\epsfig{file=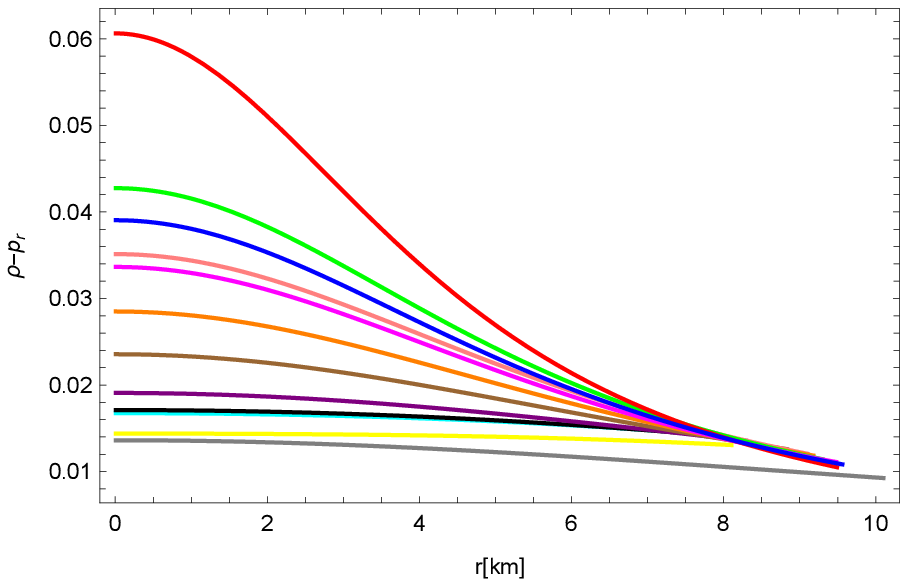,width=0.33\linewidth}  &
\epsfig{file=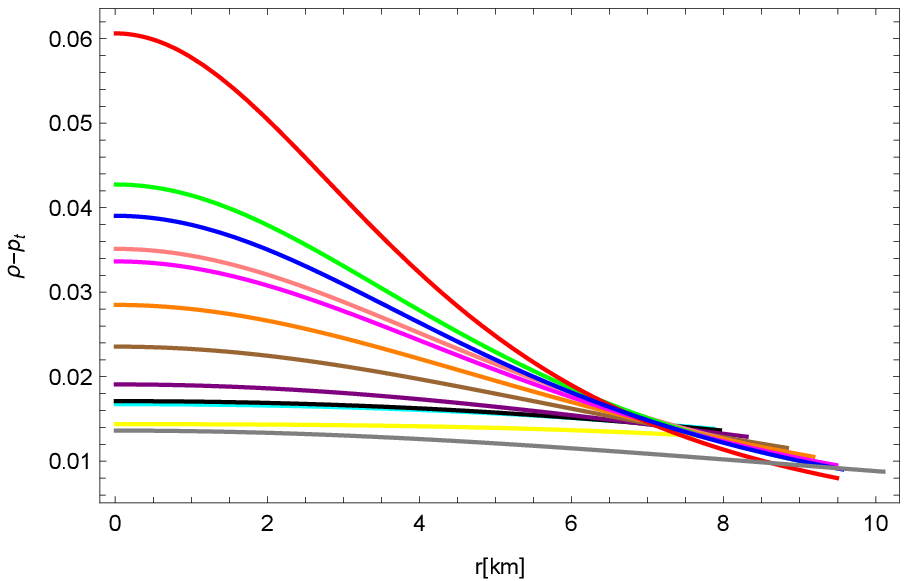,width=0.33\linewidth}  &
\epsfig{file=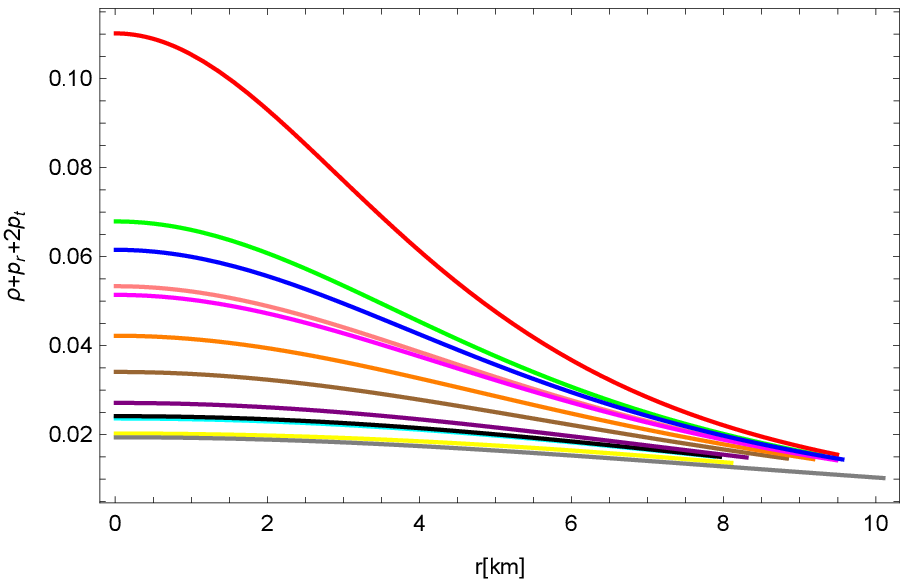,width=0.33\linewidth}  &
\end{tabular}
\caption{Profiles of the energy conditions of $S_{1}$ ({\color{cyan} $\bigstar$}), $S_{2}$ ({\color{black} $\bigstar$}), $S_{3}$ ({\color{yellow} $\bigstar$}), $S_{4}$ ({\color{violet} $\bigstar$}), $S_{5}$ ({\color{brown} $\bigstar$}), $S_{6}$ ({\color{pink} $\bigstar$}), $S_{7}$ ({\color{orange} $\bigstar$}), $S_{8}$ ({\color{green} $\bigstar$}), $S_{9}$ ({\color{magenta} $\bigstar$}), $S_{10}$ ({\color{red} $\bigstar$}), $S_{11}$ ({\color{blue} $\bigstar$}), $S_{12}$ ({\color{gray} $\bigstar$}).}\center
\label{Fig:5}
\end{figure}

\subsection{\textbf{Equilibrium Conditions}}

Now, we investigate the equilibrium condition of our model under the existence of hydrostatic force, gravitational force and anisotropic force. The equilibrium condition among all these forces can be written as
\begin{equation}
\frac{M_G(r)(\rho + p_r)}{r}e^{\frac{\lambda - \upsilon}{2}} + \frac{dp_r}{dr} - \frac{2}{r}(p_t - p_r) = 0.
\end{equation}
The effective gravitational mass $M_G(r)$ is expressed as
\begin{equation}
M_G(r) = \frac{1}{2}\upsilon^{\prime}e^{\frac{\upsilon - \lambda}{2}}.
\end{equation}
Inserting value of $M_G(r)$ in Eq. (21), we get
\begin{equation}
\frac{\upsilon^{\prime}}{2}(\rho + p_r) + \frac{dp_r}{dr} - \frac{2}{r}(p_t - p_r) = 0,
\end{equation}
where, $\mathcal{F}_g = -\frac{\upsilon^{\prime}}{2}(\rho + p_r)$, $\mathcal{F}_h = -\frac{dp_r}{dr}$ and $\mathcal{F}_a = \frac{2}{r}\Delta$ with $\mathcal{F}_g$, $\mathcal{F}_h$ and $\mathcal{F}_a$ identified as gravitational force, hydrostatic force and anisotropic force. The sum of these three forces should be equal to zero for the system to be balanced. Therefore, Eq. (23) gives
\begin{equation}
\mathcal{F}_g + \mathcal{F}_h + \mathcal{F}_a = 0.
\end{equation}
It can be easily noticed from Fig. 6, that all these forces express the required equilibrium condition.
\begin{figure} [H] \center
\begin{tabular}{cccc}
\epsfig{file=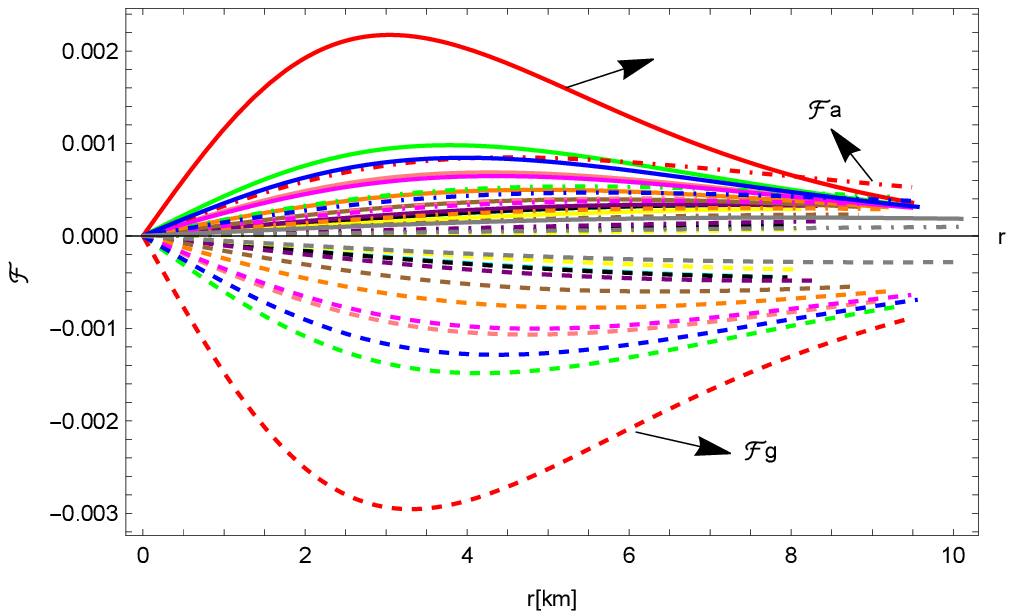,width=0.5\linewidth}  &
\end{tabular}
\caption{Behavior of $\mathcal{F}_h$, $\mathcal{F}_g$ and $\mathcal{F}_a$ of $S_{1}$ ({\color{cyan} $\bigstar$}), $S_{2}$ ({\color{black} $\bigstar$}), $S_{3}$ ({\color{yellow} $\bigstar$}), $S_{4}$ ({\color{violet} $\bigstar$}), $S_{5}$ ({\color{brown} $\bigstar$}), $S_{6}$ ({\color{pink} $\bigstar$}), $S_{7}$ ({\color{orange} $\bigstar$}), $S_{8}$ ({\color{green} $\bigstar$}), $S_{9}$ ({\color{magenta} $\bigstar$}), $S_{10}$ ({\color{red} $\bigstar$}), $S_{11}$ ({\color{blue} $\bigstar$}), $S_{12}$ ({\color{gray} $\bigstar$}).}\center
\label{Fig:6}
\end{figure}

\subsection{\textbf{Equation of State}}

Further, we determine ($EoS$) parameters for radial and tangential pressures, presented by
\begin{align}
\omega_r = \frac{p_r}{\rho}, \qquad \omega_t = \frac{p_t}{\rho}.
\end{align}
The graphical illustration of both parameters in Fig. 7, clearly shows that $0 < \omega_r$, $\omega_t < 1$. Thus, it is concluded that the $EoS$ satisfy the necessary conditions for compact stars under discussion.
\begin{figure} [H] \center
\begin{tabular}{cccc}
\epsfig{file=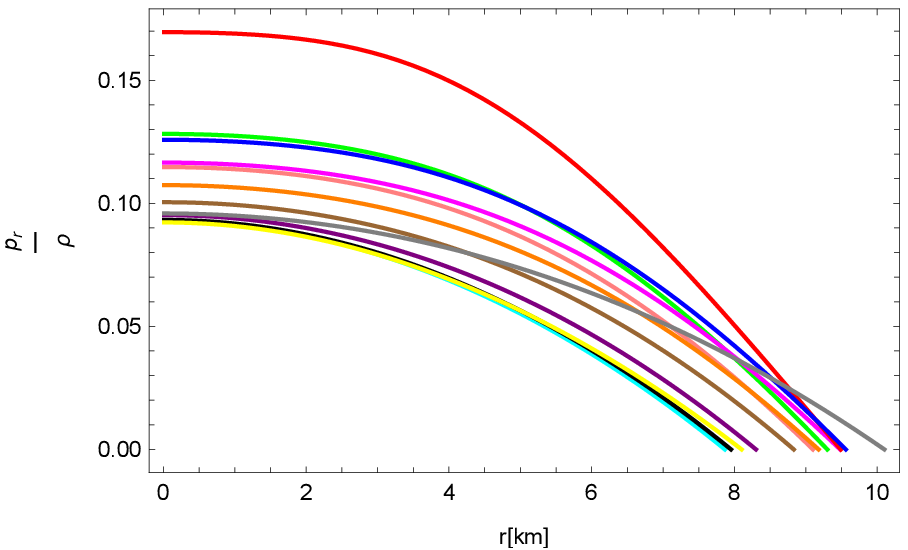,width=0.5\linewidth}  &
\epsfig{file=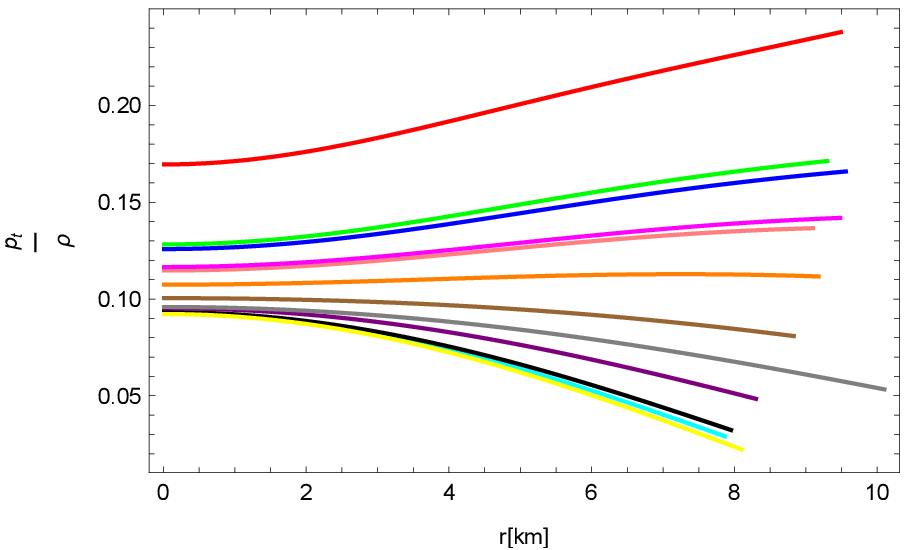,width=0.5\linewidth}  &
\end{tabular}
\caption{Profiles of $\omega_r$ and $\omega_t$ of $S_{1}$ ({\color{cyan} $\bigstar$}), $S_{2}$ ({\color{black} $\bigstar$}), $S_{3}$ ({\color{yellow} $\bigstar$}), $S_{4}$ ({\color{violet} $\bigstar$}), $S_{5}$ ({\color{brown} $\bigstar$}), $S_{6}$ ({\color{pink} $\bigstar$}), $S_{7}$ ({\color{orange} $\bigstar$}), $S_{8}$ ({\color{green} $\bigstar$}), $S_{9}$ ({\color{magenta} $\bigstar$}), $S_{10}$ ({\color{red} $\bigstar$}), $S_{11}$ ({\color{blue} $\bigstar$}), $S_{12}$ ({\color{gray} $\bigstar$}).}\center
\label{Fig:7}
\end{figure}

\subsection{\textbf{Causality Condition}}

The causality condition is subjected to the sound speed of both pressure components, i.e. radial and transverse pressures which are expressed as
\begin{align}
\upsilon_r^2 = \frac{dp_r}{d\rho}, \qquad \upsilon_t^2 = \frac{dp_t}{d\rho}.
\end{align}
Here, $\upsilon_r^2$ and $\upsilon_t^2$ stand for radial and tangential sound speed. Using the concept of Herrera \cite{Lhr} i.e., $(0 \le \upsilon_r^2, \upsilon_t^2 \le 1)$, both of the sound speeds of pressure components must vary from 0 to 1 to stabilize our system. Further, we examine the graphical response of our selected model for the Andreasson condition \cite{Han}, $|\upsilon_r^2 - \upsilon_t^2| \le 1$. The graphical representations in Fig. 8, indicate the stability and balanced nature of {$\mathcal{R}+ \alpha \mathcal{A}$ model} satisfying the causality condition.
\begin{figure} [H] \center
\begin{tabular}{cccc}
\epsfig{file=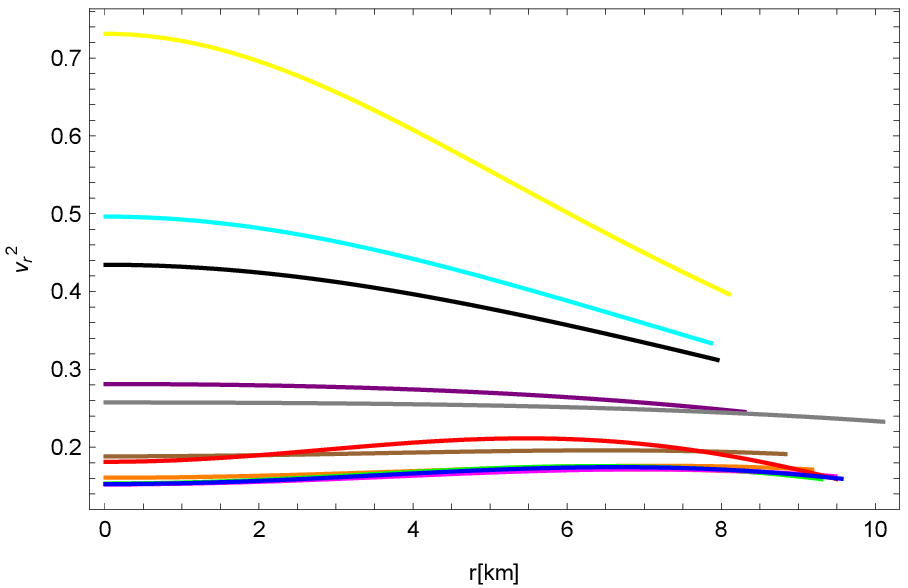,width=0.4\linewidth}  &
\epsfig{file=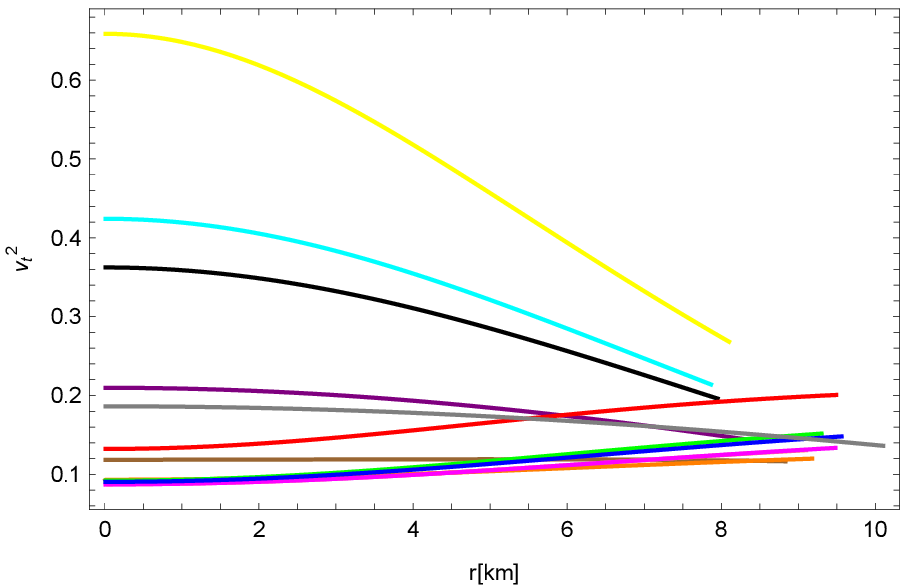,width=0.4\linewidth}  &
\end{tabular}
\begin{tabular}{cccc}
\epsfig{file=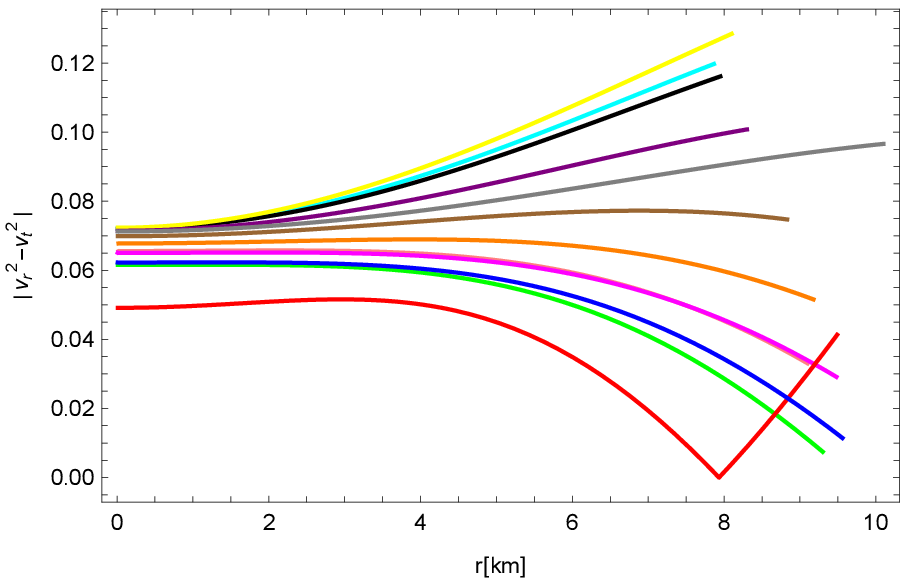,width=0.4\linewidth}  &
\epsfig{file=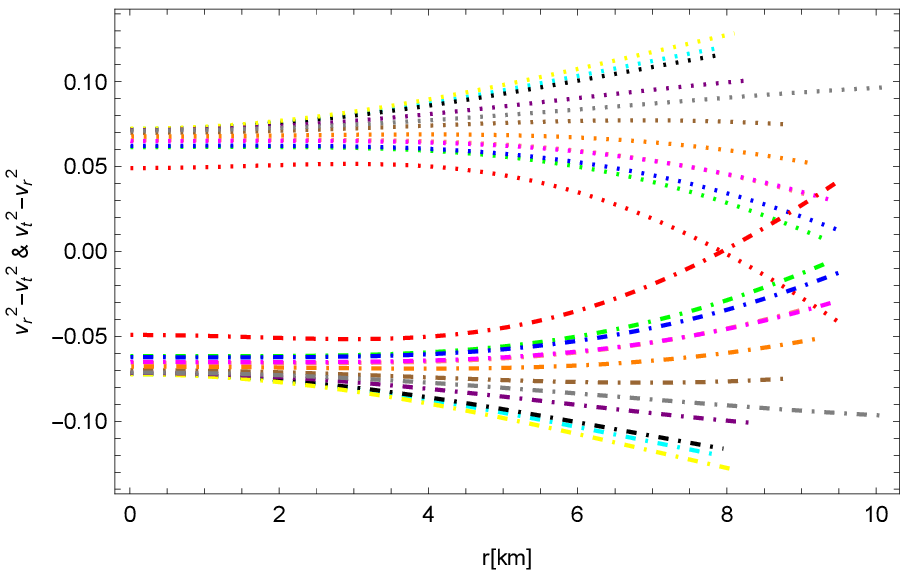,width=0.4\linewidth}  &
\end{tabular}
\caption{Behavior of speeds of sound parameters, and Abreu condition of $S_{1}$ ({\color{cyan} $\bigstar$}), $S_{2}$ ({\color{black} $\bigstar$}), $S_{3}$ ({\color{yellow} $\bigstar$}), $S_{4}$ ({\color{violet} $\bigstar$}), $S_{5}$ ({\color{brown} $\bigstar$}), $S_{6}$ ({\color{pink} $\bigstar$}), $S_{7}$ ({\color{orange} $\bigstar$}), $S_{8}$ ({\color{green} $\bigstar$}), $S_{9}$ ({\color{magenta} $\bigstar$}), $S_{10}$ ({\color{red} $\bigstar$}), $S_{11}$ ({\color{blue} $\bigstar$}), $S_{12}$ ({\color{gray} $\bigstar$}).}\center
\label{Fig:8}
\end{figure}

\subsection{\textbf{Mass Function, Compactness Factor and Redshift Analysis}}

The mass function \cite{Bhr} attained by making use of the metric potential $g_{rr}^{-} = g_{rr}^{+}$ is given as
\begin{equation}
\mathcal{M}(r) = \frac{a^2 r^3}{2[(1 + b r^2)^4 + a^2 r^2]}.
\end{equation}
Moreover, we compute the compactness parameter $\mathcal{U}(r)$ \cite{Mkm} and the redshift function $\mathcal{Z}(r)$ \cite{Cgb}, derived as the following expressions
\begin{equation}
\mathcal{U}(r) = \frac{2 \mathcal{M}(r)}{r} = \frac{a^2 r^2}{[(1 + b r^2)^4 + a^2 r^2]},
\end{equation}
\begin{equation}
\mathcal{Z}(r) = e^{-\frac{\nu}{2}} - 1.
\end{equation}
Fig. 9 shows the graphical behavior of $\mathcal{M}(r)$, $\mathcal{U}(r)$ and $\mathcal{Z}(r)$. The nature of mass function meets the Buchdahl \cite{Hab} and Bondi \cite{Hbi} constraint i.e., $\frac{2 M}{R} < \frac{8}{9}$. In addition, it can be noticed that as we advance towards the boundary, mass function and compactness parameter increase monotonically while the plot of redshift indicates monotonically decreasing behavior. Therefore, we sum up that these graphs illustrate the satisfactory behavior for our model.
\begin{figure} [H] \center
\begin{tabular}{cccc}
\epsfig{file=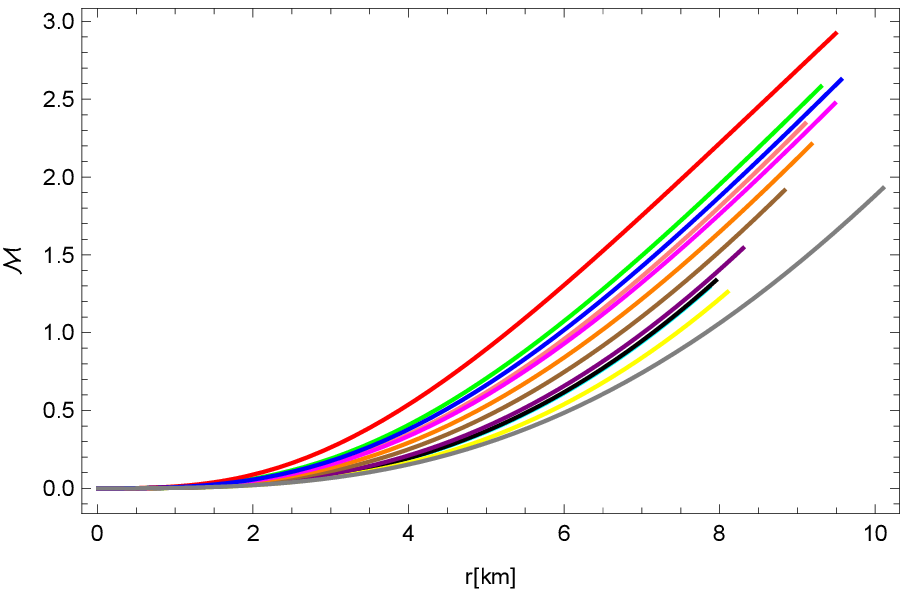,width=0.33\linewidth}  &
\epsfig{file=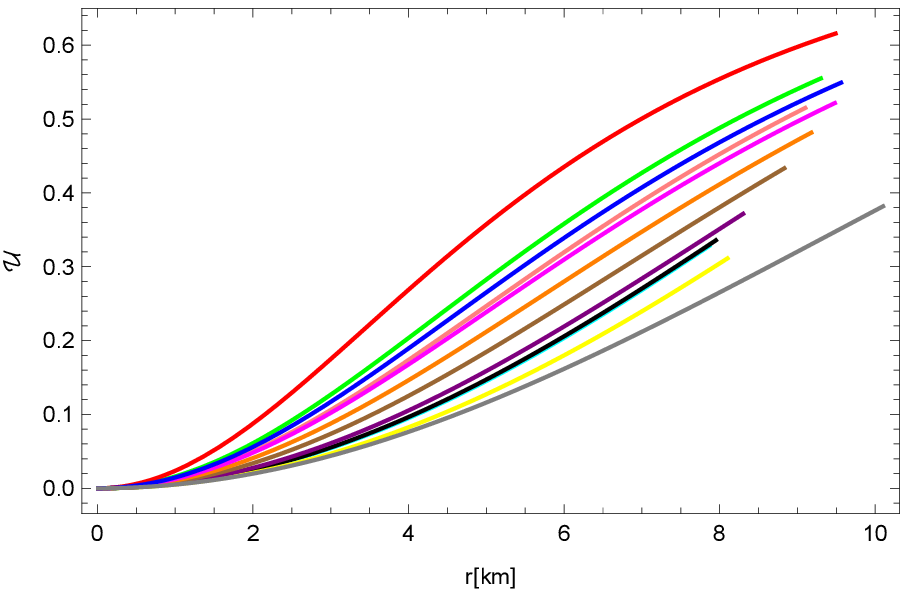,width=0.33\linewidth}  &
\epsfig{file=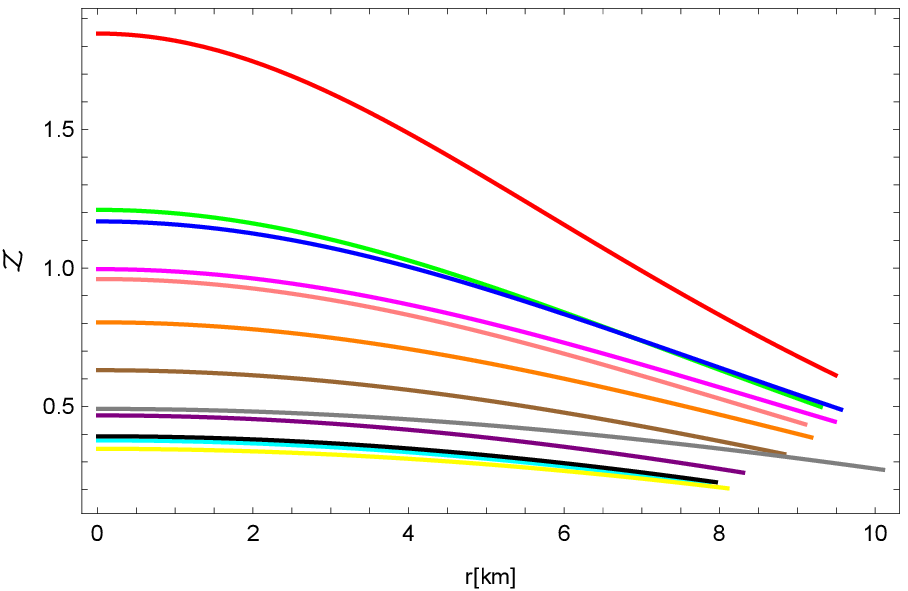,width=0.33\linewidth}  &
\end{tabular}
\caption{Graphs of mass function (left panel), compactness factor (middle panel), and redshift(right panel) of $S_{1}$ ({\color{cyan} $\bigstar$}), $S_{2}$ ({\color{black} $\bigstar$}), $S_{3}$ ({\color{yellow} $\bigstar$}), $S_{4}$ ({\color{violet} $\bigstar$}), $S_{5}$ ({\color{brown} $\bigstar$}), $S_{6}$ ({\color{pink} $\bigstar$}), $S_{7}$ ({\color{orange} $\bigstar$}), $S_{8}$ ({\color{green} $\bigstar$}), $S_{9}$ ({\color{magenta} $\bigstar$}), $S_{10}$ ({\color{red} $\bigstar$}), $S_{11}$ ({\color{blue} $\bigstar$}), $S_{12}$ ({\color{gray} $\bigstar$}).}\center
\label{Fig:9}
\end{figure}

\subsection{\textbf{Adiabatic Index}}

The stiffness of ($EoS$) for a given energy density is described by adiabatic index. It verifies the stability of stellar objects, both relativistically and non relativistically. Chandrasekhar \cite{Scd} presented the theory of dynamical stability for the infinitesimal, radial adiabatic oscillations of stellar spheres. His idea has been tested by many authors for isotropic and anisotropic stellar structures \cite{Hhn, Whi, Dho, Ddd, Hos, Iba}. The value of adiabatic index should be greater than $\frac{4}{3}$, for compact star models to be stable. Adiabatic index of pressure components, $p_r$ and $p_t$ have the following expressions respectively
\begin{align}
\gamma_r = \frac{\rho + p_r}{p_r}(\frac{dp_r}{d\rho}) = \frac{\rho + p_r}{p_r} v_r^2, \qquad \gamma_t = \frac{\rho + p_t}{p_t}(\frac{dp_t}{d\rho}) = \frac{\rho + p_t}{p_t} v_t^2.
\end{align}
It is concluded from Fig. 10 that the considered {$\mathcal{R}+ \alpha \mathcal{A}$ model} is potentially balanced under the above stability condition.
\begin{figure} [H] \center
\begin{tabular}{cccc}
\epsfig{file=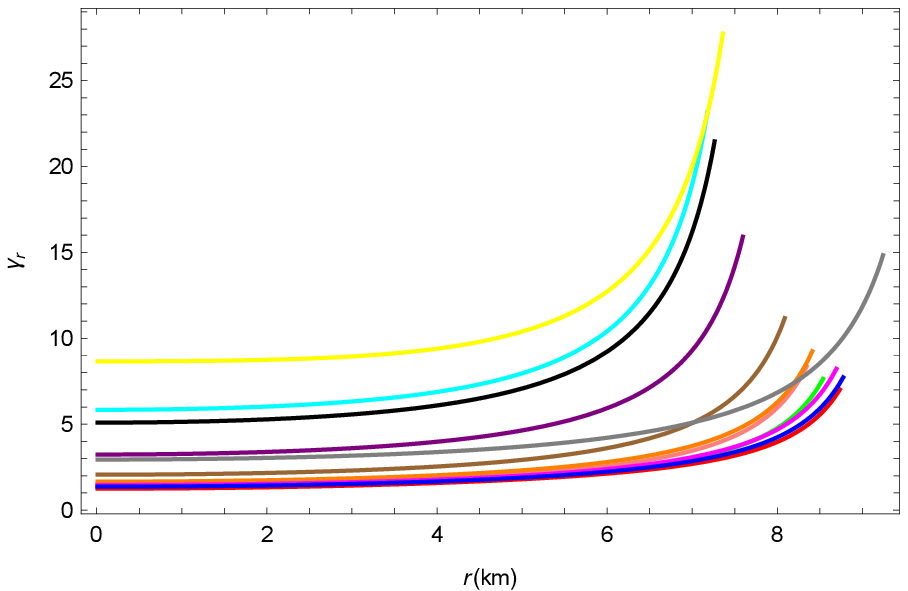,width=0.5\linewidth}  &
\epsfig{file=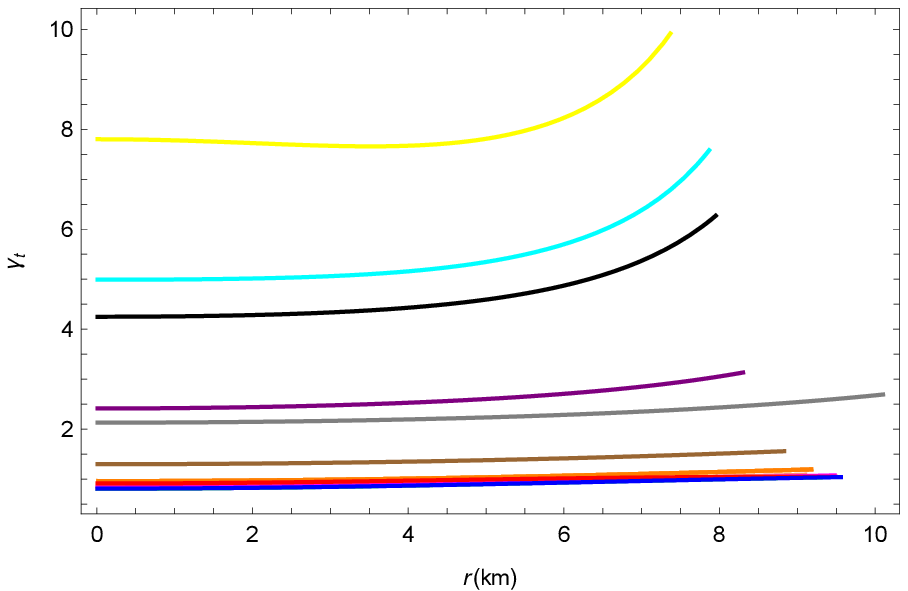,width=0.5\linewidth}  &
\end{tabular}
\caption{Variation of $\gamma_r$, and $\gamma_t$ of $S_{1}$ ({\color{cyan} $\bigstar$}), $S_{2}$ ({\color{black} $\bigstar$}), $S_{3}$ ({\color{yellow} $\bigstar$}), $S_{4}$ ({\color{violet} $\bigstar$}), $S_{5}$ ({\color{brown} $\bigstar$}), $S_{6}$ ({\color{pink} $\bigstar$}), $S_{7}$ ({\color{orange} $\bigstar$}), $S_{8}$ ({\color{green} $\bigstar$}), $S_{9}$ ({\color{magenta} $\bigstar$}), $S_{10}$ ({\color{red} $\bigstar$}), $S_{11}$ ({\color{blue} $\bigstar$}), $S_{12}$ ({\color{gray} $\bigstar$}).}\center
\label{Fig:10}
\end{figure}

\section{Summary and Concluding Remarks}

We assume a realistic {$\mathcal{R}+ \alpha \mathcal{A}$ gravity model} by using the Karmarkar condition to investigate a novel family of embedded class-I solutions in the anisotropic framework. Since the Karmarkar condition provides a link between $g_{rr}$ and $g_{tt}$, it simplifies the solution generating process of $\mathcal{EFE}$ to a single metric potential. Assuming the metric potential $e^{\lambda} = 1 + \frac{a^{2} r^{2}}{(1 + b r^{2})^{4}}$ for this purpose, we use the Karmarkar condition to determine the second metric potential, $e^{\nu} = \left[\left(A - \frac{a B}{2 b (1 + b r^{2})}\right)^{2}\right]$, where $a$, $b$, $A$, and $B$ are arbitrary constants. We also determine the unknown constants by applying matching constraints between the internal and Schwarzschild external geometries. We examine the physical behavior of twelve different compact stars to confirm that the discussed solutions are physically feasible.
The primary objective of our research is to study stellar structures using a viable {$\mathcal{R}+ \alpha \mathcal{A}$ gravity model} in the context of an anisotropic matter source. The notable results are listed below.\\\\
\begin{itemize}
\item{Metric potentials are frequently used to characterize the properties of spacetime. According to Fig. 1, both the metric potentials $g_{rr} = e^{\lambda}$ and $g_{tt} = e^{\nu}$ exhibit graphical behavior that is finite, positive, singularity-free, and satisfies the criteria, i.e., $e^{\lambda(r=0)} = 1$ and $e^{\nu(r=0)} = \left[A - \frac{a B}{2 b} \right]^{2}$. Since these graphs expand monotonically and reach their maximum values at the boundary, the progression of both potentials exhibits excellent outcomes, supporting the {$\mathcal{R}+ \alpha \mathcal{A}$ gravity model}.}
\item{The trend of density and pressure components for {$\mathcal{R}+ \alpha \mathcal{A}$ gravity model} is shown in Fig. 2. {For this purpose, we consider the model parameter $\alpha = 1.3\times10^{-11}$. It is important to mention here that the parameter $\alpha$ has a very limited range which provides us physically accepted results. We have checked two other values of parameter $\alpha = 1.0\times10^{-9}$, $\alpha = 1.9\times10^{-8}$ and found that there is a very slight difference in the behavior of the graphs towards the boundary. That difference is shown for one star only $Vela~X-1$ as a magnified image in Fig. 2.} The variations of $\rho$, $p_r$, and $p_t$ with respect to $r$ are finite and regular at the center. It is also evident that graphs of $\rho$, $p_r$, and $p_t$ reach their highest values at the center and show a diminishing response toward the border, supporting the stability of our model.
The gradient of $\rho$, $p_r$, and $p_t$ has been depicted in Fig. 3. These graphs exhibit negative behavior, demonstrating the consistency of our suggested model.}
\item{It is clear from Fig. 4 that anisotropy is positive throughout the stellar objects. This behavior demonstrates that anisotropy has a repulsive nature, confirming its existence for compact objects.}
\item{The energy bounds for the suggested {$\mathcal{R}+ \alpha \mathcal{A}$ model} are shown in Fig. 5. We observed that our given model fulfills all energy bounds.}
\item{The equilibrium conditions of $\mathcal{F}_g$, $\mathcal{F}_h$, and $\mathcal{F}_a$ of our physically acceptable model show balancing behavior in Fig. 6.}
\item{The graphical response of $EoS$ must fulfill the requirements $0 < \omega_r$, $\omega_t < 1$ for the consistency of the stellar objects. The corresponding graphical representation in Fig. 7 depicts that both $EoS$ ratios have a stable character.}
\item{The radial component sound velocity $\upsilon_{r}^{2}$ plot and the transverse component sound velocity $\upsilon_{t}^{2}$ plot should fall within the range [0, 1] for compact stars. According to Fig. 8, the stability constraints for our model are stable.}
\item{From Fig. 9, it is easy to notice that the graphs of the mass function and the compactness parameter demonstrate monotonically increasing responses. On the other hand, the graphical illustration of redshift analysis reveals monotonically decreasing behavior. The behavior shows that our given model achieves the stability criteria.}
\item{In Fig. 10, the adiabatic index of $\gamma_r$ and $\gamma_t$ for our given model is more than $\frac{4}{3}$, confirming the stability of our system.\\\\}
\end{itemize}
As a final observation, we have used the Karmarkar condition to provide a stellar system that is both well-stable and singularity-free. We conclude that {$\mathcal{R}+ \alpha \mathcal{A}$ gravity model} supports the existence of compact objects which follow observable patterns. {Furthermore, it is worth mentioning that our suggested results are almost similar to the outcomes investigated by Naz et al. \cite{Naz} in the context of $f(\mathcal{R})$ theory of gravity. It may be noted that due to the involvement of inverse curvature terms, we get more dense stars in our case.}

\section*{Acknowledgements}
The authors appreciate important comments from anonymous reviewer which improved the paper.

\newpage
\section*{Appendix}
Following are the values of $\Gamma_{i}$’s,
\begin{align*}
\Gamma_{1} = -\frac{a (1 + b r^2) (2 a^2 r^2 - (-3 + b r^2) (1 + b r^2)^3) B (-2 b (1 + b r^2) A + a B)}{2 b (a^2 r^2 + (1 + b r^2)^4)^2}
\end{align*}
\begin{align*}
\Gamma_{2} = \frac{(2 a (2 a^3 b r^2 (1 + b r^2) A - 2 a b (1 + b r^2)^4 (-3 + 5 b r^2) A - a^4 r^2 B + a^2 (-3 + b r^2) (1 + b r^2)^3 B + 2 b (-3 + b r^2) (1 + b r^2)^6 B))}{((a^2 r^2 + (1 + b r^2)^4)^2 (-2 b (1 + b r^2) A + a B))}
\end{align*}
\begin{align*}
\Gamma_{3} = -\frac{2 b (a^2 r^2 + (1 + b r^2)^4)^2}{a (1 + b r^2) (2 a^2 r^2 - (-3 + b r^2) (1 + b r^2)^3) B (-2 b (1 + b r^2) A + a B)}
\end{align*}
\begin{align*}
\Gamma_{4}& = \frac{(a^2 r^2 + (1 + b r^2)^4)^2 (2 b (1 + b r^2) A - a B)}{2 a b (1 + b r^2)^3 (2 a^2 r^2 - (-3 + b r^2) (1 + b r^2)^3) B} \\& + \frac{(a^2 r^2 + (1 + b r^2)^4)^2 (-2 b (1 + b r^2) A + a B)}{2 a (1 + b r^2)^3 (-1 + 3 b r^2) (-2 a b (1 + b r^2) A + a^2 B + b (1 + b r^2)^3 B)} \\& + \frac{2 (a^2 r^2 + (1 + b r^2)^4)^2 (2 b (1 + b r^2) A - a B)}{a (-2 a^3 b r^2 (1 + b r^2) A + 4 a b (-1 + b r^2) (1 + b r^2)^4 A + a^4 r^2 B + 2 a^2 (1 + b r^2)^3 B + 2 b (1 + b r^2)^7 B)}
\end{align*}
\begin{align*}
\Gamma_{5} = -\frac{(a^2 r^2 + (1 + b r^2)^4)^4 (2 b (1 + b r^2) A - a B)}{a^2 (1 + b r^2)^4 (2 a^2 r^2 - (-3 + b r^2) (1 + b r^2)^3)^2 B^2 (-2 b (1 + b r^2) A + a B)}
\end{align*}
\begin{align*}
\Gamma_{5}^{\prime} = -\frac{8 r (a^2 r^2 + (1 + b r^2)^4)^3 (a^4 r^2 (-1 + b r^2) + b (-5 + b r^2) (1 + b r^2)^7 - a^2 (1 + b r^2)^3 (2 + b r^2 (-1 + 7 b r^2)))}{a^2 (1 + b r^2)^5 (2 a^2 r^2 - (-3 + b r^2) (1 + b r^2)^3)^3 B^2}
\end{align*}
\begin{align*}
\Gamma_{5}^{\prime\prime}& = (8 (a^2 r^2 + (1 + b r^2)^4)^2 (2 a^8 r^6 (3 + b r^2 (-12 + 5 b r^2)) + b (1 + b r^2)^{14} (15 + b r^2 (181 + b r^2 (-67 + 7 b r^2)))\\& - a^6 r^4 (1 + b r^2)^3 (-25 + b r^2 (93 + b r^2 (-151 + 35 b r^2))) - 2 a^2 (1 + b r^2)^{10} (-3 + b r^2 (-37 + b r^2 (-71 + b r^2\\& (-211 + 34 b r^2)))) + a^4 r^2 (1 + b r^2)^6 (31 + 3 b r^2 (-23 + b r^2 (125 + b r^2 (-13 + 122 b r^2)))))) /\\& (a^2 (1 + b r^2)^6 (-2 a^2 r^2 + (-3 + b r^2) (1 + b r^2)^3)^4 B^2)
\end{align*}
\begin{align*}
\Gamma_{6} = \frac{(a^2 r^2 + (1 + b r^2)^4)^3 (2 b (1 + b r^2) A - a B) (-2 b (1 + b r^2) A + a B)}{4 a^2 (1 + b r^2)^2 (-1 + 3 b r^2)^2 (-2 a b (1 + b r^2) A + a^2 B + b (1 + b r^2)^3 B)^2}
\end{align*}
\begin{align*}
\Gamma_{6}^{\prime}& = (r (a^2 r^2 + (1 + b r^2)^4)^2 (-48 b^4 (-1 + b r^2) (1 + b r^2)^9 A^2 B - 6 a^5 b (3 + b r^2 + b^2 r^4 + 3 b^3 r^6) A B^2 + a^6 (3 + b r^2 (-2 + 3 b r^2)) B^3\\& + a^4 b (1 + b r^2)^2 B (12 b (3 + b r^2 (-2 + 3 b r^2)) A^2 - (-1 + b r^2) (1 + b r^2) (19 + 3 b r^2) B^2) + 4 a b^3 (1 + b r^2)^7 A (16 b (-2 + 3 b r^2) A^2\\& + (1 + b r^2) (-11 + 9 b r^2) B^2) + 4 b^2 (a + a b r^2)^3 A (-2 b (3 + b r^2 (-2 + 3 b r^2)) A^2 + (1 + b r^2) (-27 + b r^2 (19 + 18 b r^2)) B^2)\\& + 2 a^2 b^2 (1 + b r^2)^5 B (-6 b (-17 + b r^2 (10 + 19 b r^2)) A^2 - (-5 + 3 b r^2) (B + b r^2 B)^2))) / (2 a^2 (1 + b r^2)^3 (-1 + 3 b r^2)^3\\& (-2 a b (1 + b r^2) A + a^2 B + b (1 + b r^2)^3 B)^3)
\end{align*}
\begin{align*}
\Gamma_{6}^{\prime\prime}& = -(((a^2 r^2 + (1 + b r^2)^4) (48 b^5 (1 + b r^2)^{16} (1 + b r^2 (23 + b r^2 (-37 + 21 b r^2))) A^2 B^2 - 8 a^9 b r^2 (1 + b r^2) (15 + \\&b r^2 (-8 + b r^2 (94 + 3 b r^2 (-16 + 9 b r^2)))) A B^3 + a^{10} r^2 (15 + b r^2 (-8 + b r^2 (94 + 3 b r^2 (-16 + 9 b r^2)))) B^4\\& + 4 a b^4 (1 + b r^2)^{14} A B (-8 b (7 + 25 b r^2 (7 + 3 b r^2 (-5 + 3 b r^2))) A^2 - (1 + b r^2)(11 + 3 b r^2 (79 - 93 b r^2 + 45 b^2 r^4)) B^2)\\& + 4 a^3 b^3 (1 + b r^2)^{10} A B (-4 b (35 + b r^2 (1100 + 3 b r^2 (-618 + b r^2 (-28 + 561 b r^2)))) A^2 - (1 + b r^2) (43 + b r^2 (1042 +\\& b r^2 (-700 - 546 b r^2 + 513 b^2 r^4))) B^2) + 4 a^7 b (1 + b r^2)^3 A B (-8 b^2 r^2 (15 + b r^2 (-8 + b r^2 (94 +\\& 3 b r^2 (-16 + 9 b r^2)))) A^2 - (1 + b r^2) (6 + b r^2 (355 + 3 b r^2 (4 + b r^2 (98 + 39 b r^2 (-2 + 3 b r^2))))) B^2) + 4 a^5 b^2 \\&(1 + b r^2)^6 A B (-4 b (6 + b r^2 (325 + b r^2 (106 + 3 b r^2 (-56 + b r^2 (-8 + 105 b r^2))))) A^2 - (1 + b r^2) (41 + b r^2 (1376 + \\&b r^2 (-871 + 9 b r^2 (-31 + 15 b r^2 (6 + 5 b r^2))))) B^2) + a^8 (B + b r^2 B)^2 (24 b^2 r^2 (15 + b r^2 (-8 + b r^2 (94 + 3 b r^2 \\&(-16 + 9 b r^2)))) A^2 + (1 + b r^2) (3 + b r^2 (185 + b r^2 (-24 + b r^2 (280 + 3 b r^2 (-49 + 45 b r^2))))) B^2) + 2 a^2 b^3 (1 + b r^2)^{12}\\& (64 b^2 (2 + b r^2 (59 + 3 b r^2 (-56 + 45 b r^2))) A^4 + 2 b (85 + b r^2 (2152 + 3 b r^2 (-650 - 480 b r^2 + 603 b^2 r^4)))\\& A^2 B^2 + (1 + b r^2)^2 (5 + b r^2 (101 + 3 b r^2 (-23 + 9 b r^2))) B^4) + a^4 b^2 (1 + b r^2)^8 (16 b^2 (1 + b r^2) (3 + b r^2 \\&(149 + 3 b r^2 (-77 + 45 b r^2))) A^4 + 4 b (114 + b r^2 (3773 + b r^2 (-2344 + 3 b r^2 (-1310 + 9 b r^2 (154 + 127 b r^2))))) A^2 B^2\\& + (1 + b r^2)^2 (29 + b r^2 (660 + b r^2 (-242 + 3 b r^2 (-100 + 63 b r^2)))) B^4) + a^6 b (1 + b r^2)^4 (16 b^3 r^2 (15 + b r^2 \\&(-8 + b r^2 (94 + 3 b r^2 (-16 + 9 b r^2)))) A^4 + 8 b (1 + b r^2) (9 + b r^2 (510 + b r^2 (95 + 3 b r^2 (25 + b r^2 (-68 + 177 b r^2)))))\\& A^2 B^2 + (1 + b r^2)^2 (22 + b r^2 (745 + 3 b r^2 (-174 + b r^2 (212 - 60 b r^2 + 81 b^2 r^4)))) B^4)))/(2 a^2 (1 - 3 b r^2)^4 \\&(1 + b r^2)^4 (-2 a b (1 + b r^2) A + a^2 B + b (1 + b r^2)^3 B)^4))
\end{align*}
\begin{align*}
\Gamma_{7} = \frac{(a^2 r^2 + (1 + b r^2)^4)^4 (2 b (1 + b r^2) A - a B) (-2 b (1 + b r^2) A + a B)}{a^2 r^2 (-2 a^3 b r^2 (1 + b r^2) A + 4 a b (-1 + b r^2) (1 + b r^2)^4 A + a^4 r^2 B + 2 a^2 (1 + b r^2)^3 B + 2 b (1 + b r^2)^7 B)^2}
\end{align*}
\begin{align*}
\Gamma_{7}^{\prime}& = (2 (a^2 r^2 + (1 + b r^2)^4)^3 (-8 b^3 (1 + b r^2)^{12} (-1 + 3 b r^2) A^2 B + 6 a^7 b r^4 (1 + b r^2) A B^2 - a^8 r^4 B^3 + 2 a^2 b (1 + b r^2)^8 B (12 b \\&(1 + b r^2 (-12 + 7 b r^2)) A^2 - (-1 + b r^2) (B + b r^2 B)^2) + 8 a b^2 (1 + b r^2)^9 A (-2 b (1 + b r^2 (-12 + 7 b r^2)) A^2 + (-1 + 2 b r^2) (B +  \\&b r^2 B)^2) + 2 (a + a b r^2)^4 B (-2 b^2 r^2 (9 + b r^2 (8 + 63 b r^2)) A^2 + (-1 + b r^2) (-1 + 11 b r^2) (B + b r^2 B)^2) + 4 a^3 b (1 + b r^2)^5 A \\&(2 b^2 r^2 (3 + b r^2 (2 + 23 b r^2)) A^2 - (3 + b r^2 (-36 + 25 b r^2)) (B + b r^2 B)^2) - a^6 (r + b r^3)^2 B (3 B^2 + b r^2 (12 b A^2 + (4 + 13 b \\&r^2) B^2)) + 2 a^5 b r^2 (1 + b r^2)^3 A (9 B^2 + b r^2 (4 b A^2 + (10 + 53 b r^2) B^2)))) / (a^2 r^3 (-2 a^3 b r^2 (1 + b r^2) A + 4 a b (-1 + b r^2) \\&(1 + b r^2)^4 A + a^4 r^2 B + 2 a^2 (1 + b r^2)^3 B + 2 b (1 + b r^2)^7 B)^3)
\end{align*}
\begin{align*}
\Gamma_{7}^{\prime\prime}& = (2 (a^2 r^2 + (1 + b r^2)^4)^2 (-48 b^4 (1 + b r^2)^{22} (1 + b r^2 (-2 + 5 b r^2)) A^2 B^2 + 8 a^{13} b r^8 (1 + b r^2) A B^3 - a^{14} r^8 B^4 + 2 a^{12} r^6 (1 + b r^2)\\& B^2 (-12 b^2  r^2 (1 + b r^2) A^2 - (3 + b r^2 (7 + 5 b r^2 (7 + 11 b r^2))) B^2) + 8 a^9 b (r + b r^3)^4 A B (4 b^2 r^2 (6 + b r^2 (11 + 3 b r^2 (38 + 67 b r^2)))\\& A^2 + (1 + b r^2) (17 + b r^2 (16 + b r^2 (564 + b r^2 (270 + 257 b r^2)))) B^2) + 4 a^8 b r^4 (1 + b r^2)^5 (-8 b^3 r^2 (3 + b r^2 (5 + b r^2 (61 \\&+ 107 b r^2))) A^4 - 2 b (1 + b r^2) (51 + b r^2 (36 + b r^2 (1942 + 3 b r^2 (348 + 533 b r^2)))) A^2 B^2 + (1 + b r^2)^3 (-37 + b r^2 (-48 + b r^2 (-85 \\&+ 358 b r^2))) B^4) + a^{10} r^4 (1 + b r^2)^3 (-16 b^4 r^4 (1 + b r^2) A^4 - 16 b^2 r^2 (9 + b r^2 (18 + 7 b r^2 (22 + 39 b r^2))) A^2 B^2 + (1 \\&+ b r^2) (-17 + b r^2 (-20 + b r^2 (-462 + b r^2 (-92 + 31 b r^2)))) B^4) + 16 a b^3 (1 + b r^2)^{19} A B (4 b (3 + b r^2 (-21 + b r^2 (43 - 29 b r^2)))\\& A^2 + 3 (1 + b r^2 (-1 + 2 b r^2)) (B + b r^2 B)^2) + 8 b^2 r^4 (a + a b r^2)^7 A B (4 b (17 + b r^2 (8 + b r^2 (712 + b r^2 (356 + 755 b r^2)))) \\&A^2 - (-1 + b r^2) (145 + b r^2 (448 + 1367 b r^2)) (B + b r^2 B)^2) + 16 a^3 b^2 (1 + b r^2)^{15} A B (12 b (2 + b r^2 (-23 + b r^2 (186 + b r^2 (-195 \\&+ 58 b r^2)))) A^2 + (9 + b r^2 (-51 + b r^2 (-3 + b r^2))) (B + b r^2 B)^2) + 16 a^5 b (1 + b r^2)^{11} A B (2 b^3 r^4 (139 + b r^2 (455 + b r^2 (1381\\& - 983 b r^2))) A^2 + (6 + b r^2 (-63 + b r^2 (511 + b r^2 (-853 + 231 b r^2)))) (B + b r^2 B)^2) - 4 a^2 b^2 (1 + b r^2)^{16} (16 b^2 (3 + b r^2 (-36 \\&+ b r^2 (290 + b r^2 (-300 + 91 b r^2)))) A^4 - 24 b (1 + b r^2)^2 (-3 + b r^2 (19 + b r^2 (-21 + 13 b r^2))) A^2 B^2 + (3 + b^2 r^4) (B \\&+ b r^2 B)^4) + 8 a^4 b (1 + b r^2)^{12} (16 b^4 r^4 (-17 + b r^2 (-62 + b r^2 (-173 + 112 b r^2))) A^4 - 2 b (1 + b r^2)^2 (18 + b r^2 (-198 + b r^2 \\&(1605 + b r^2 (-1988 + 543 b r^2)))) A^2 B^2 - (3 + b r^2 (-15 + b r^2 (-23 + 19 b r^2))) (B + b r^2 B)^4) + 4 a^6 (1 + b r^2)^8 (-4 b^4 r^4 (17 +\\&b r^2 (4 + b r^2 (758 + 324 b r^2 + 897 b^2 r^4))) A^4 + 4 b^3 r^4 (1 + b r^2)^2 (-213 + b r^2 (-586 + 3 b r^2 (-623 + 584 b r^2))) A^2 B^2 - (3 \\&+ b r^2 (-30 + b r^2 (243 + 4 b r^2 (-147 + 50 b r^2)))) (B + b r^2 B)^4) + 8 a^{11} b r^6 (1 + b r^2)^2 A B (6 B^2 + b r^2 (4 b (1 + b r^2) A^2 +\\& (13 + b r^2 (88 + 153 b r^2)) B^2)))) / (a^2 r^4 (-2 a^3 b r^2 (1 + b r^2) A + 4 a b (-1 + b r^2) (1 + b r^2)^4 A + a^4 r^2 B + 2 a^2 (1 + b r^2)^3 B\\& + 2 b (1 + b r^2)^7 B)^4)
\end{align*}
\begin{align*}
\Gamma_{8} = \frac{2 a (-1 + 3 b r^2) (-2 a b (1 + b r^2) A + a^2 B + b (1 + b r^2)^3 B)}{(1 + b r^2) (a^2 r^2 + (1 + b r^2)^4) (2 b (1 + b r^2) A - a B)}
\end{align*}
\begin{align*}
\Gamma_{9} = \frac{(1 + b r^2) (a^2 r^2 + (1 + b r^2)^4) (2 b (1 + b r^2) A - a B)}{2 a (-1 + 3 b r^2) (-2 a b (1 + b r^2) A + a^2 B + b (1 + b r^2)^3 B)}
\end{align*}
\begin{align*}
\Gamma_{10} = \frac{a r^2 (-2 a^3 b r^2 (1 + b r^2) A + 4 a b (-1 + b r^2) (1 + b r^2)^4 A + a^4 r^2 B + 2 a^2 (1 + b r^2)^3 B + 2 b (1 + b r^2)^7 B)}{(a^2 r^2 + (1 + b r^2)^4)^2 (-2 b (1 + b r^2) A + a B)}
\end{align*}
\begin{align*}
\Gamma_{11} = \frac{(a^2 r^2 + (1 + b r^2)^4)^2 (-2 b (1 + b r^2) A + a B)}{a r^2 (-2 a^3 b r^2 (1 + b r^2) A + 4 a b (-1 + b r^2) (1 + b r^2)^4 A + a^4 r^2 B + 2 a^2 (1 + b r^2)^3 B + 2 b (1 + b r^2)^7 B)}
\end{align*}

\end{document}